\documentclass[letterpaper, 12pt, parskip=full,DIV=10]{scrartcl}
\usepackage{subfiles}
\usepackage{packages}
\newcommand{\blind}{0}
\usepackage[super, sort]{natbib} 
\title{Battling the Coronavirus `Infodemic' Among Social Media Users in {Kenya and Nigeria}}
\author[1]{Molly Offer-Westort*}
\author[2]{Leah R. Rosenzweig}
\author[3]{Susan Athey}
\affil[1]{Department of Political Science, University of Chicago, Chicago, IL, USA; mollyow@uchicago.edu}
\affil[2]{Development Innovation Lab, University of Chicago, Chicago, IL, USA}
\affil[3]{Stanford Graduate School of Business, Stanford University, Stanford, CA, USA}
\date{\today}

\makeatletter
\let\Title\@title
\makeatother

\begin{document}%

\def\spacingset#1{\renewcommand{\baselinestretch}%
{#1}\small\normalsize} \spacingset{1}

\if0\blind
{
  \maketitle
} \fi

\if1\blind
{
  \bigskip
  \bigskip
  \bigskip
  \begin{center}
    {\LARGE\textbf{\Title}}
\end{center}
  \medskip
} \fi

\bigskip


{\textbf{Abstract:}}
\begin{abstract} 

\small
How can we induce social media users to be discerning when sharing information during a pandemic? An experiment on Facebook Messenger with users from Kenya {($n$ = 7,498)} and Nigeria {($n$ = 7,794)} tested interventions designed to decrease intentions to share COVID-19 misinformation without decreasing intentions to share factual posts. The initial stage of the study incorporated: (i) a factorial design with 40 intervention combinations; and (ii) a contextual adaptive design, increasing the probability of assignment to treatments that worked better for previous subjects with similar characteristics. The second stage evaluated the best-performing treatments and a targeted treatment assignment policy estimated from the data. We precisely estimate null effects from warning flags and related article suggestions, tactics used by social media platforms. 
However, nudges to consider information's accuracy reduced misinformation sharing {relative to control} by 4.9\% {(estimate = $-2.3$~pp, s.e. = $1.0$, $Z = -2.31$, $p = 0.021$, 95\% CI = [$-4.2$, $-0.35$])}. Such low-cost scalable interventions may improve the quality of information circulating online.

\end{abstract}


\clearpage



Amid the outbreak of the novel coronavirus (SARS-CoV-2), people around the world were also subjected to an ``infodemic''—the spread of misinformation related to the virus.  
This study evaluates interventions designed to deter users from sharing misinformation on social media without adversely affecting how they share true information. {We operationalize this objective with a composite outcome, sharing discernment, that gives false information twice as much (negative) weight as sharing true information.} Our application focuses on information and misinformation about prevention and treatment for COVID-19; we began the study in February 2021 before vaccines for the virus were widely available. Using targeted Facebook advertisements, we recruited a sample of social media users in Kenya and Nigeria, two of the three largest Facebook markets in sub-Saharan Africa \citep{world-population-reviewfacebook}. Users interacted with a Facebook Messenger chatbot to answer survey questions and receive randomized treatments, keeping users who selected into the study on the platform where they already engage with similar media posts.

{Our experiment proceeded in two stages. %
The first (learning) stage implemented a design to learn the best-performing interventions from 40 factorial combinations of treatments. We used an adaptive treatment assignment algorithm that, over time, gave higher assignment probabilities to treatments that had performed well for previous participants with similar characteristics.}  Adaptive designs have two benefits: first, they improve outcomes for individuals during the experiment, as compared to uniform random assignment. Improving participants' outcomes is important when conducting research on a topic like misinformation, where the literature has noted concerns of unintended negative consequences  \citep{swire2020searching}.  
Second, adaptive designs can lead to a more effective treatment ``policy'' learned at the end of the experiment \citep{even2006action, caria2020adaptive, kasy2021adaptive, athey2022contextual}. {In this setting, a policy is a decision rule that assigns treatments to users on the basis of their characteristics. Adaptive designs can improve the quality of learned policies}
 because they allocate more individuals to conditions that perform well, {allowing us to more precisely differentiate among the best candidate treatments}; if there are many poorly performing treatments, this advantage is larger. 

{Each user's treatment consisted of two independently randomized factors: (i)} (headline-level) treatments delivered on specific posts shown to treated users{, such as} flags or warning labels pinned on the article of interest; {(ii)} (respondent-level) messaging treatments delivered to treated users{, such as} tips for spotting fake news, training videos, and nudges. These two types of interventions vary in their cost and scalability. While headline flags require fact checking sources to keep up with the generation of new misinformation, {educational content to combat misinformation sharing can be used over a longer period and across contexts.}
 
The second (evaluation) stage estimated treatment effects for the most effective interventions identified during the learning stage, comparing them against each other and to control. We recruited a new sample of Facebook users and randomly assigned them to{: (i) a pure control condition; (ii) a condition where all false posts they saw were accompanied with fact check labels; (iii) a condition where all false posts they saw were accompanied with articles on related topics from verified sources; (iv) a condition that nudged respondents to be more attentive to the accuracy of posts \citep{pennycook2021shifting}; (v) a condition sharing  Facebook's tips for spotting misinformation \citep{guessetal2020digital}; and (vi)} a ``Learned Targeted Policy'' that assigned different respondent-level treatments based on user characteristics. 

{
This study evaluates whether these interventions increase discernment, that is, whether they decrease sharing of false COVID-19 cures relative to any reduction in sharing of true information. We focus on sharing rather than belief as our main outcome of interest because even if an individual does not believe misinformation, sharing exposes others to misinformation and can have harmful effects.  For example, in Iran, dozens of people died from alcohol poisoning after ingesting methanol supposedly due to the rumor that alcohol could prevent coronavirus \citep{haghdoost2020alcohol}. The interventions we selected for study are drawn from the academic literature as well as from interventions that have been used by industry. As the interventions operate through different mechanisms, comparing their performance informs the debate about whether misinformation spreads due to inattention or people do not have skills to spot it \citep{ecker2022psychological}. In addition, as most of the existing experimental research focuses on the Global North, our study brings comparative data to a global problem. Two recent exceptions from sub-Saharan Africa include a field experiment in Zimbabwe using Whatsapp messages from a trusted NGO to counter COVID misinformation \citep{bowles2020center} and a recent survey among traders in Lagos, Nigeria looking at the correlates of belief in COVID-related misinformation \citep{Grossman2020}. Although we might expect interventions that rely on psychological techniques to operate similarly in diverse contexts, evidence suggests that some interventions found to be effective in the US, Canada, and Europe produce varying results in India \citep{badrinathan2021educative, guessetal2020digital} and Pakistan \citep{ali2021countering} (see ref. \citenum{roozenbeek2022countering}). One reason for this may be varying levels of digital literacy. This study explores this and other potential sources of heterogeneity.
}


\section{Results}
\paragraph{Study sample}
We recruited Facebook users 18 years and older in Kenya and Nigeria through targeted Facebook advertisements \citep{Rosenzweig_2020}. 
The learning and evaluation stages had 4,761 {(Kenya: $n = 2,180$; Nigeria: $n = 2,581$)} and 10,531 {(Kenya: $n = 5,318$; Nigeria: $n = 5,213$)} distinct participants, respectively. We learned which treatments were most effective in the learning stage and separately obtained precise estimates of treatment effects using evaluation stage data. 
In Supplementary~\autoref*{appendix:demographics} we compare sample characteristics with nationally representative Afrobarometer surveys{; our samples' characteristics were roughly similar to the Afrobarometer estimates in terms of gender and age, although slightly younger and more educated}. 

\paragraph{Survey experience}\label{section:experience}
{Users who clicked on our advertisements were prompted to start a conversation with our research page's Messenger chatbot. Once we obtained informed consent, users were asked a series of questions.} 
{Prior to treatment, we showed participants four media posts (two true and two false in random order) drawn from our stimuli set. 
After viewing each stimulus, users were first asked whether they wanted to share the post (publicly) on their Facebook Timeline, and then asked whether they wanted to share it (privately) through Messenger. 
In the next portion of the survey, users were asked about news and media consumption; this section was designed as a ``distractor'' module between pretest measurement and assignment of treatment.}

{At this point, randomization occurred; covariates and the pretest response measurement were sent to our algorithm, treatment assignment probabilities were calculated, and treatment was assigned according to those probabilities and returned. In the learning stage, the design was factorial: users were assigned to both a headline-level and a respondent-level condition. In the evaluation stage, users were assigned either type of treatment, but not both, to facilitate more straightforward estimation.}

{For users assigned a respondent-level treatment, we next delivered messaging associated with the treatment.  
All users were then shown four new stimuli (two true and two false), e.g., each user was shown a total of eight unique posts during the survey. If the user was assigned a headline-level treatment, this treatment was applied only to the posttreatment misinformation stimuli. Here, our design was intended to align with social media platform practice, as flags and fact checking labels are not generally applied to true posts. 
For each stimuli we asked the same self-reported sharing intention questions. } 

\paragraph{Primary outcomes}
{To construct our preregistered combined response measure of sharing discernment, 
}
{we code responses to each of the questions following stimuli as one if the user affirmed they wanted to share the post and zero otherwise. Let $M_i$ be the sum of respondent $i$'s posttest responses to the misinformation stimuli and let $T_i$ be the equivalent for the true informational stimuli. 
Intentions to share false stimuli are given a weight of $-1$ and intentions to share true stimuli are given a weight of 0.5 in this measure, to reflect the idea that sharing misinformation is potentially more harmful than the benefit of sharing true information:} 
\begin{equation}
Y_i = -M_i + 0.5 T_i
\end{equation}

{We consider implications of alternative weighting schemes in \autoref*{fig:alternative_weighting}.} We also report results disaggregated by type of stimuli (true or false) and sharing channel.  
{When analyzing impact of treatments, we use a pretest-posttest design \citep{broockman2017design}, estimating treatment effects adjusting for users' pretreatment sharing intentions.} {\autoref*{tab:alternative_results} 
evaluates alternative specifications and illustrates that controlling for pretest responses reduces the standard error of estimates by about 20\%.}


\paragraph{Learning stage}
We designed the learning stage to discover which treatment conditions were most effective at increasing sharing discernment.  We used a factorial design: one factor with seven respondent-level interventions and one factor with four headline-level interventions, alongside a baseline control condition for each. 
\autoref*{tab:treatments} describes all of the interventions tested. 

{To assign treatment in the learning stage, we used an adaptive assignment algorithm, a version of balanced linear Thompson sampling (BLTS) \citep{dimakopoulou2017estimation,dimakopoulou2019balanced}. 
Our contextual Thompson sampling algorithm proceeds in several steps. We start with a prior about the distribution of the response measure under each treatment arm for each possible realization of individual characteristics (the ``context''). 
As in linear Thompson sampling (LinTS), the algorithm estimates a model that allows the outcome to be a linear function of covariates, {in our case,} including interactions of the covariates with treatment indicators. {Our outcome model is regularized only on covariates and treatment covariate interactions, with no penalty applied to main effects.}
As we collect data, the algorithm calculates and updates the Bayesian posterior distributions. Using a batched approach, participants are grouped according to their order of arrival, and this updating occurs at the end of each batch. 
Treatment assignment probabilities for each context are then set as equal to the posterior probability that each treatment has the highest posterior mean. 
In balanced linear Thompson sampling, we reweight the data using inverse probability weights, to account for biases that might otherwise arise due to nonuniform assignment in previous batches.}

{We select this algorithm type as ref. \citenum{dimakopoulou2017estimation} demonstrate that BLTS has a faster rate of learning than LinTS and a linear upper confidence bound algorithm (LinUCB) in a range of simulated settings, particularly when misspecification of the outcome model may be a concern. {We perform simulations using a model fit to the first non-adaptive batch of our learning stage data, and find performance benefits of BLTS which are directionally consistent with those of \citenum{dimakopoulou2017estimation}.
However, we caveat that there may be variation in performance of different assignment algorithms across different settings, and so there may be cases in which LinTS or LinUCB outperform BLTS.} In addition, algorithms that assign treatments with known and nonzero probabilities (such as Thompson Sampling, but not LinUCB) have advantages when using the data to test hypotheses, estimate the impact of counterfactual policies, or estimate optimal treatment assignment policies using data from the learning stage \citep{zhan2022policy}.} We further incorporate a lower bound on assignment probabilities to improve policy learning at the end of the learning stage \citep{zhan2022policy}. 

This adaptive design allowed us to continue to learn which treatments were best, while reducing the probability that users were assigned to ineffective or harmful interventions. 
\autoref*{tab:learning_cumulative} illustrates how the probability that users will be assigned to our targeted policies of interest, the Learned Targeted Policy and the Restricted Targeted Policy, increases over successive batches of the adaptive experiment. 

In the learning stage, average response on sharing discernment is $-0.445$ (s.e. = 0.020); we estimate average response under uniform random assignment would be $-0.456$ (s.e = 0.027) using adaptively weighted estimators \citep{zhan2021off}. While the difference is small and not statistically significant, the higher value under our algorithm indicates that the algorithm directionally improved mean within-experiment response over what response would have been had we used uniform assignment in the learning stage, equivalent to a 1\% reduction in misinformation sharing (difference estimate = $0.011$, s.e. = $0.033$, $Z$ = 0.327, p = 0.743, 95\% CI = [$-0.054$, $0.076$]). The relatively small improvement may be due to misassignment early in the adaptive experiment, as the algorithm may initially follow false leads before gathering more data and updating the response model. 

\begin{figure}[H]
\centering
  \includegraphics[width=\textwidth]{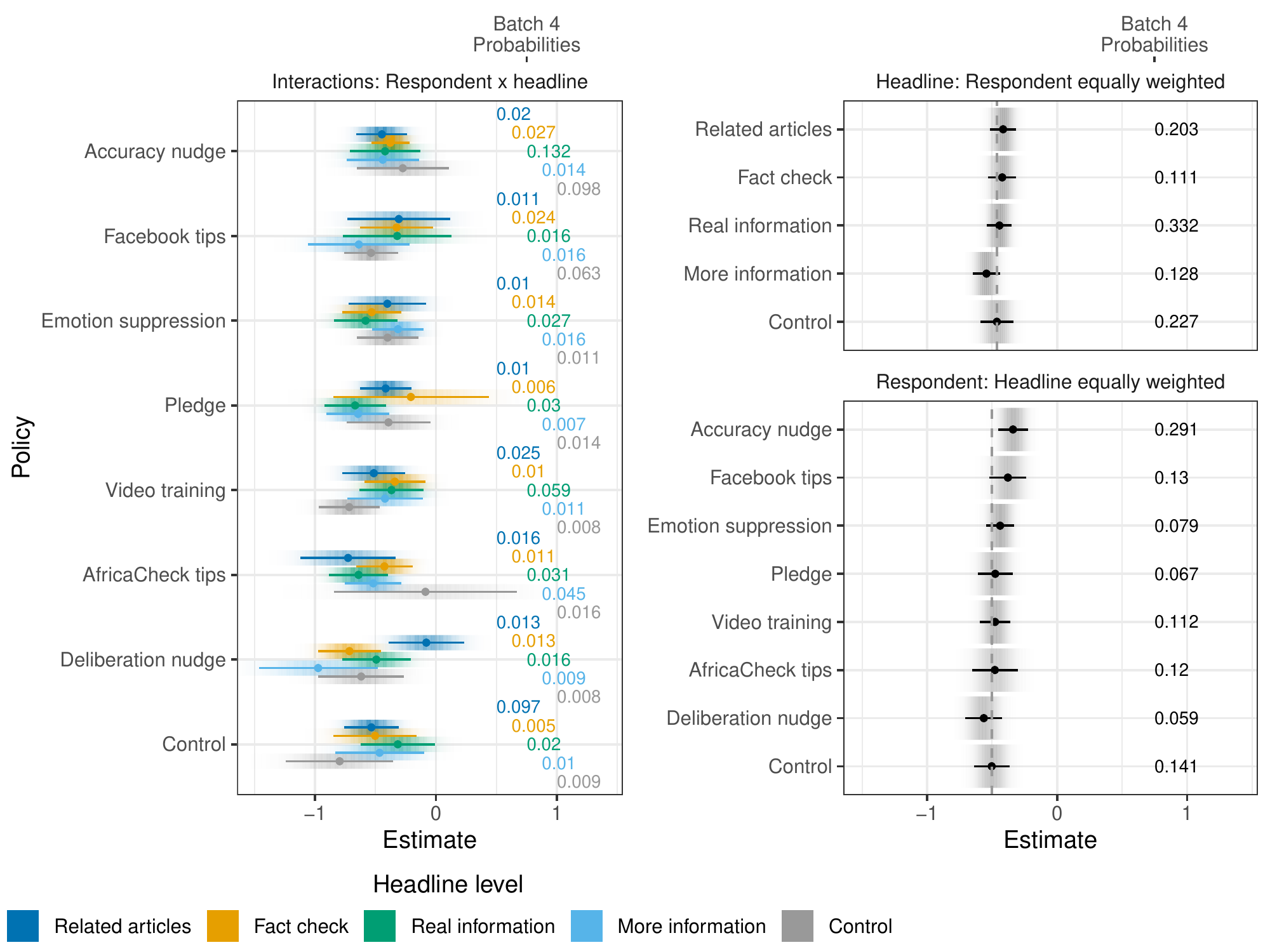}
   \caption{{\textbf{Learning stage estimates.} The sample is users in the learning stage, total $n = 4,761$.    
   Estimates are of mean response in terms of sharing discernment; error bars represent 95\% confidence intervals. Panels represent estimates under different types of policies: (1) unique respondent (row) $\times$ headline (color) interactions; (2) headline factor levels with all respondent levels assigned with equal probability; (3) respondent factor levels with all headline levels assigned with equal probability. Estimates are produced from an augmented inverse probability weighted estimator, as described in Section~\ref{section:estimation}, using contextual adaptive weights as described by ref. \citenum{zhan2021off} to achieve valid inference under the contextual adaptive algorithm. Adaptive weights for each panel are calculated separately, and so estimates in panels (2) and (3) are not direct averages across estimates in the first panel. Batch 4 probabilities are average assignment probabilities within a condition in the final batch of the learning stage of the study. }
}   \label{fig:learning_estimates}

\end{figure}

\autoref{fig:learning_estimates} reports estimated response under different treatment combinations in the learning stage. 
{We allowed for the possibility of interactions between the headline- and respondent-level treatments, as illustrated in the left panel of \autoref{fig:learning_estimates}. 
However, our experiment is not well-powered to detect and compare these interaction effects. Our 40 unique treatment combinations have as few as 89 respondents in each condition, and the possibility for discovering spurious effects is high. 
Estimates of differences of respondent- and headline-interacted response from the headline average and respondent averages are presented in \autoref*{tab:learning_interaction_headline} and \autoref*{tab:learning_interaction_respondent}.}

Our objective in the learning stage was to learn the treatments associated with the highest estimated mean sharing discernment in each factor{, illustrated in the right panels of \autoref{fig:learning_estimates}}. 
These treatments were the accuracy nudge and Facebook tips (respondent-level) and fact check and related articles (headline-level); examples of each are presented in \autoref{section:methods}. 

{Because our design targets learning about what works best, by gathering data in the evaluation stage about the most promising treatments, we can not draw strong conclusions about whether treatments not selected for evaluation are better or worse than the control. 
}

\paragraph{Evaluation stage}
{We designed the evaluation stage to obtain precise estimates of the treatment effects (relative to control) of the best treatments from the learning stage.} Treatment was assigned with equal probability to each of the two most effective treatments from each factor (respondent- and headline-level) from the learning stage 
(accuracy nudge, Facebook tips, fact check, and related articles), 
or to the Learned Targeted Policy, which assigned users, based on their characteristics, to one of four respondent-level treatments.

{To construct the Learned Targeted Policy, we fit a generalized random forest model \citep{athey2016generalized} to the learning stage data, using sharing discernment as the outcome. We use this model to predict what mean discernment would be under each treatment, conditional on a covariate profile. The policy's decision rule is to assign to each user in the evaluation stage the treatment associated with the highest estimated mean outcome, conditional on their covariates. 
We focus only on respondent-level treatments to yield a better comparison between the best overall treatments (accuracy nudge, Facebook tips{, as discussed below}) and a personalized targeted policy.
}

{We also consider the Restricted Targeted Policy under which users could only be assigned to the accuracy nudge or Facebook tips. The Restricted Targeted Policy is constructed by first estimating, with learning stage data, the difference in outcomes between the two treatments using a causal forest. Motivated by our observation that treatment effect heterogeneity is largely associated with the false sharing outcome (\autoref*{tab:heterogeneity_treatment}), we use false sharing as the outcome here. Although the Restricted Targeted Policy was not directly assigned during the evaluation stage, because unconfoundedness holds in the evaluation stage (conditional on characteristics, the treatment assignment probabilities are known and based only on observables), and overlap also holds (conditional on characteristics, there is a positive probability that each respondent-level treatment is assigned) we can estimate the (counterfactual) value of the Restricted Targeted Policy. We use standard techniques for evaluation of policies that are distinct from those used in the data collection process, using augmented inverse propensity weighted estimators \citep{robins1995semiparametric}, as described in \autoref{section:estimation}. We note that estimation of the Restricted Targeted Policy was not preregistered.}

{
If we applied the Learned Targeted Policy to all participants in the evaluation stage, we would assign \num{83.1}\% the accuracy nudge, \num{15.6}\% Facebook tips, and the rest either the emotion suppression or video treatment. If we applied the Restricted Targeted Policy, we would assign \num{78.9}\% the accuracy nudge and the rest Facebook tips. For \num{73.6}\% of the participants in the evaluation stage, the Learned Targeted Policy and the Restricted Targeted Policy are aligned. }

\paragraph{Systematic differences in sharing under control}
{Before analyzing the impact of our interventions, we present initial results about baseline levels of sharing intentions. }
Under the control condition, users have systematically different sharing intentions for false posts relative to true posts, both overall and across sharing channels. Users report intentions to share true stimuli at rates 45.6\% higher than false stimuli on any channel (estimate = $21.3$~pp, s.e. = $0.9$, $Z = 22.91$, $p < 0.001$, 95\% CI = [$19.45$, $23.09$]).  Users also intend to share true stimuli on their Timeline at rates 4.8\% higher compared to privately on Messenger (estimate = $2.8$~pp, s.e. = $0.7$, $Z = 3.79$, $p < 0.001$, 95\% CI = [$1.36$, $4.28$]), but at rates 7.5\% lower for false stimuli (estimate = $-3.1$~pp, s.e. = $0.6$, $Z = -4.91$, $p < 0.001$, 95\% CI = [$-4.4$, $-1.89$]). It seems participants can differentiate between true and false posts to some extent and are more reluctant to share false posts in a more public forum. 

We preregistered several participant characteristics of interest for analysis of treatment effect heterogeneity, including variables collected by social media platforms (age, gender) and variables of theoretical interest (political allegiance, digital literacy, scientific knowledge). Here, we examine the effects of these variables under the control condition to illuminate which participants are most likely to share more misinformation.

Our results suggest that younger users, men, those aligned with the ruling party, participants with low digital literacy, and those with low scientific knowledge have relatively lower sharing discernment in the control condition (see \autoref{tab:heterogeneity_control}). The differences are large in magnitude, e.g., men are $18.9$\% more likely to share false posts than women {(estimate = $-8.0$~pp, s.e. = $1.7$, $Z = -4.76$, $p < 0.001$, 95\% CI = [$-11.31$, $-4.72$])}, while low digital literacy users are $24.3$\% more likely to share false posts {(estimate = $10.1$~pp, s.e. = $1.7$, $Z = 5.98$, $p < 0.001$, 95\% CI = [$6.76$, $13.35$])}.

\begin{figure}[H]
\centering
\includegraphics[width = \textwidth]{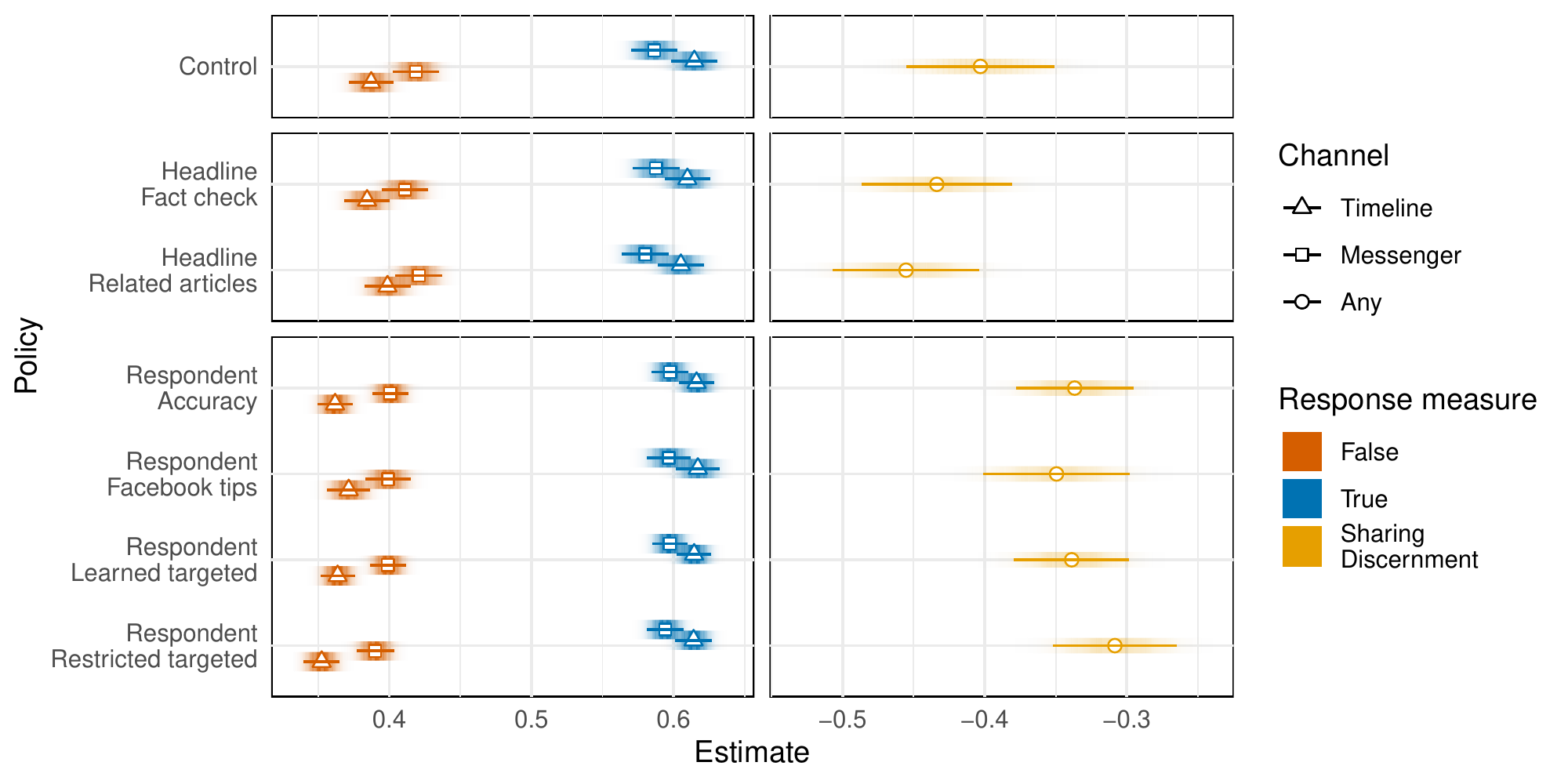}
\caption{\textbf{Response estimates.} The sample is users in the evaluation stage, $n = 10,531$. Response measures are average propensity to share false and true posts over either channel, and a sharing discernment measure. Estimates are of mean response, and are produced from an augmented inverse probability weighted estimator, as described in \autoref{section:estimation}. Error bars represent 95\% confidence intervals.}
\label{fig:main_results}

\end{figure}

\begin{table}[H]
   \centering
\resizebox{\textwidth}{!}
{
\begin{tabular}[t]{lccccccc}
 & \textbf{Sharing} &  & \textbf{False} &  &  & \textbf{True} & \\
 & \textbf{Discernment} & Any sharing & Messenger & Timeline & Any sharing & Messenger & Timeline\\\cmidrule(lr){2-2} \cmidrule(lr){3-5} \cmidrule(lr){6-8} \multicolumn{4}{l}{\textbf{Age}} \rule{0pt}{1.2\normalbaselineskip}\\
\hspace{1em} Below median & \num{-0.488} & \num{0.481} & \num{0.436} & \num{0.391} & \num{0.668} & \num{0.578} & \num{0.590}\\
\hspace{2em}(n = 5,300) & (\num{0.036}) & (\num{0.012}) & (\num{0.012}) & (\num{0.011}) & (\num{0.011}) & (\num{0.012}) & (\num{0.012})\\
\hspace{1em} Above median & \num{-0.316} & \num{0.451} & \num{0.401} & \num{0.383} & \num{0.689} & \num{0.595} & \num{0.640}\\
\hspace{2em}(n = 5,231) & (\num{0.040}) & (\num{0.012}) & (\num{0.012}) & (\num{0.011}) & (\num{0.011}) & (\num{0.012}) & (\num{0.012})\\\cmidrule(lr){2-8}
\hspace{1em} Difference & \num{-0.172} & \num{0.030} & \num{0.035} & \num{0.008} & \num{-0.021} & \num{-0.017} & \num{-0.050}\\
\hspace{2em} & (\num{0.053}) & (\num{0.017}) & (\num{0.017}) & (\num{0.016}) & (\num{0.016}) & (\num{0.017}) & (\num{0.017})\\
 & $p =$ \num{0.001} & $p =$ \num{0.075} & $p =$ \num{0.036} & $p =$ \num{0.622} & $p =$ \num{0.186} & $p =$ \num{0.302} & $p =$ \num{0.003}\\
 & {}[\num{-0.276}, \num{-0.068}] & {}[\num{-0.003}, \num{0.063}] & {}[\num{0.002}, \num{0.067}] & {}[\num{-0.023}, \num{0.039}] & {}[\num{-0.053}, \num{0.010}] & {}[\num{-0.050}, \num{0.016}] & {}[\num{-0.083}, \num{-0.017}]\\\multicolumn{4}{l}{\textbf{Gender}} \rule{0pt}{1.2\normalbaselineskip}\\
\hspace{1em} Not male & \num{-0.346} & \num{0.423} & \num{0.381} & \num{0.343} & \num{0.639} & \num{0.534} & \num{0.563}\\
\hspace{2em}(n = 4,915) & (\num{0.037}) & (\num{0.012}) & (\num{0.012}) & (\num{0.011}) & (\num{0.012}) & (\num{0.012}) & (\num{0.012})\\
\hspace{1em} Male & \num{-0.453} & \num{0.503} & \num{0.452} & \num{0.426} & \num{0.713} & \num{0.632} & \num{0.660}\\
\hspace{2em}(n = 5,616) & (\num{0.038}) & (\num{0.012}) & (\num{0.011}) & (\num{0.011}) & (\num{0.011}) & (\num{0.011}) & (\num{0.011})\\\cmidrule(lr){2-8}
\hspace{1em} Difference & \num{0.107} & \num{-0.080} & \num{-0.071} & \num{-0.082} & \num{-0.075} & \num{-0.098} & \num{-0.096}\\
\hspace{2em} & (\num{0.053}) & (\num{0.017}) & (\num{0.017}) & (\num{0.016}) & (\num{0.016}) & (\num{0.017}) & (\num{0.017})\\
 & $p =$ \num{0.043} & $p =$ \num{<0.001} & $p =$ \num{<0.001} & $p =$ \num{<0.001} & $p =$ \num{<0.001} & $p =$ \num{<0.001} & $p =$ \num{<0.001}\\
 & {}[\num{0.003}, \num{0.211}] & {}[\num{-0.113}, \num{-0.047}] & {}[\num{-0.103}, \num{-0.038}] & {}[\num{-0.113}, \num{-0.051}] & {}[\num{-0.106}, \num{-0.043}] & {}[\num{-0.131}, \num{-0.065}] & {}[\num{-0.129}, \num{-0.063}]\\\multicolumn{4}{l}{\textbf{Supports governing party}} \rule{0pt}{1.2\normalbaselineskip}\\
\hspace{1em} Not aligned & \num{-0.346} & \num{0.439} & \num{0.390} & \num{0.357} & \num{0.659} & \num{0.563} & \num{0.587}\\
\hspace{2em}(n = 7,360) & (\num{0.032}) & (\num{0.010}) & (\num{0.010}) & (\num{0.010}) & (\num{0.010}) & (\num{0.010}) & (\num{0.010})\\
\hspace{1em} Aligned & \num{-0.535} & \num{0.529} & \num{0.486} & \num{0.457} & \num{0.723} & \num{0.641} & \num{0.679}\\
\hspace{2em}(n = 3,171) & (\num{0.049}) & (\num{0.015}) & (\num{0.015}) & (\num{0.014}) & (\num{0.014}) & (\num{0.015}) & (\num{0.014})\\\cmidrule(lr){2-8}
\hspace{1em} Difference & \num{0.189} & \num{-0.090} & \num{-0.096} & \num{-0.099} & \num{-0.064} & \num{-0.079} & \num{-0.092}\\
\hspace{2em} & (\num{0.058}) & (\num{0.018}) & (\num{0.018}) & (\num{0.017}) & (\num{0.017}) & (\num{0.018}) & (\num{0.018})\\
 & $p =$ \num{0.001} & $p =$ \num{<0.001} & $p =$ \num{<0.001} & $p =$ \num{<0.001} & $p =$ \num{<0.001} & $p =$ \num{<0.001} & $p =$ \num{<0.001}\\
 & {}[\num{0.075}, \num{0.303}] & {}[\num{-0.126}, \num{-0.054}] & {}[\num{-0.131}, \num{-0.062}] & {}[\num{-0.133}, \num{-0.065}] & {}[\num{-0.098}, \num{-0.030}] & {}[\num{-0.114}, \num{-0.044}] & {}[\num{-0.127}, \num{-0.058}]\\\multicolumn{4}{l}{\textbf{Digital literacy index}} \rule{0pt}{1.2\normalbaselineskip}\\
\hspace{1em} Below median & \num{-0.544} & \num{0.515} & \num{0.469} & \num{0.438} & \num{0.696} & \num{0.606} & \num{0.637}\\
\hspace{2em}(n = 5,418) & (\num{0.038}) & (\num{0.012}) & (\num{0.012}) & (\num{0.011}) & (\num{0.011}) & (\num{0.011}) & (\num{0.011})\\
\hspace{1em} Above median & \num{-0.254} & \num{0.414} & \num{0.365} & \num{0.334} & \num{0.660} & \num{0.566} & \num{0.590}\\
\hspace{2em}(n = 5,113) & (\num{0.038}) & (\num{0.012}) & (\num{0.012}) & (\num{0.011}) & (\num{0.012}) & (\num{0.012}) & (\num{0.012})\\\cmidrule(lr){2-8}
\hspace{1em}  Difference & \num{-0.290} & \num{0.101} & \num{0.104} & \num{0.104} & \num{0.035} & \num{0.039} & \num{0.047}\\
\hspace{2em} & (\num{0.053}) & (\num{0.017}) & (\num{0.017}) & (\num{0.016}) & (\num{0.016}) & (\num{0.017}) & (\num{0.017})\\
 & $p =$ \num{<0.001} & $p =$ \num{<0.001} & $p =$ \num{<0.001} & $p =$ \num{<0.001} & $p =$ \num{0.028} & $p =$ \num{0.019} & $p =$ \num{0.005}\\
 & {}[\num{-0.394}, \num{-0.186}] & {}[\num{0.068}, \num{0.133}] & {}[\num{0.072}, \num{0.137}] & {}[\num{0.072}, \num{0.135}] & {}[\num{0.004}, \num{0.067}] & {}[\num{0.007}, \num{0.072}] & {}[\num{0.014}, \num{0.080}]\\\multicolumn{4}{l}{\textbf{Scientific knowledge index}} \rule{0pt}{1.2\normalbaselineskip}\\
\hspace{1em} Below median & \num{-0.451} & \num{0.482} & \num{0.437} & \num{0.401} & \num{0.685} & \num{0.587} & \num{0.619}\\
\hspace{2em}(n = 5,560) & (\num{0.037}) & (\num{0.012}) & (\num{0.011}) & (\num{0.011}) & (\num{0.011}) & (\num{0.012}) & (\num{0.012})\\
\hspace{1em} Above median & \num{-0.349} & \num{0.448} & \num{0.398} & \num{0.372} & \num{0.671} & \num{0.586} & \num{0.610}\\
\hspace{2em}(n = 4,971) & (\num{0.039}) & (\num{0.012}) & (\num{0.012}) & (\num{0.012}) & (\num{0.011}) & (\num{0.012}) & (\num{0.012})\\\cmidrule(lr){2-8}
\hspace{1em} Difference & \num{-0.102} & \num{0.035} & \num{0.039} & \num{0.029} & \num{0.014} & \num{0.001} & \num{0.009}\\
\hspace{2em} & (\num{0.053}) & (\num{0.017}) & (\num{0.017}) & (\num{0.016}) & (\num{0.016}) & (\num{0.017}) & (\num{0.017})\\
 & $p =$ \num{0.056} & $p =$ \num{0.040} & $p =$ \num{0.019} & $p =$ \num{0.067} & $p =$ \num{0.373} & $p =$ \num{0.948} & $p =$ \num{0.607}\\
 & {}[\num{-0.206}, \num{0.003}] & {}[\num{0.002}, \num{0.068}] & {}[\num{0.006}, \num{0.071}] & {}[\num{-0.002}, \num{0.061}] & {}[\num{-0.017}, \num{0.046}] & {}[\num{-0.032}, \num{0.034}] & {}[\num{-0.024}, \num{0.041}]\\
\end{tabular}
}
   \caption{\footnotesize\textbf{Heterogeneity in response under the control condition by selected covariates.} The sample is users in the evaluation stage, $n = 10,531$. 
   Columns denote response measures, which include discernment, a weighted sum of number of false sharing intentions (negatively weighted) and true sharing intentions (positively weighted); and for false and true posts separately, average propensity to share posts over any channel, over Messenger only, and on Timeline only. 
   Estimates are of mean response under the control condition and are produced from an augmented inverse probability weighted estimator, as described in \autoref{section:estimation}, within specified subgroups. 
   {Standard errors are presented beneath estimates in parentheses; for differences, two-sided p-values and confidence intervals follow.}
 }
\label{tab:heterogeneity_control}

\end{table}

\clearpage

\paragraph{Main treatment effects}
{\autoref{fig:main_results} and} \autoref{tab:main_results} show our prespecified comparisons of each evaluated treatment condition against the control. 
The two headline-level treatments are not effective at improving sharing discernment. 
The fact check treatment is associated with a decrease, {not statistically distinguishable from zero}, of 0.6~pp (s.e. = $1.1$, $Z = -0.51$, $p = 0.608$, 95\% CI = [$-2.78$, $1.63$]) in false sharing intentions as compared to control; the effect would need to be nearly four times as large with the same degree of uncertainty for the confidence interval to exclude zero. The related articles treatment increases intention to share false stimuli as compared to control, although this estimate is not statistically distinguishable from zero (estimate = $0.8$~pp, s.e. = $1.1$, $Z = 0.71$, $p = 0.477$, 95\% CI = [$-1.39$, $2.98$]).  
{These results are not dissimilar to Facebook's own findings, observing backfire effects of such headline-level interventions \citep{TIME_facebook}.}

{The accuracy nudge treatment, however, was effective.} The accuracy nudge and Facebook tips {are associated with} increase{s in} sharing discernment of 0.066 (s.e. = $0.032$, $Z = 2.074$, $p = 0.038$, 95\% CI = [$0.004$, $0.129$]) and 0.054 (s.e. = $0.036$, $Z = 1.501$, $p = 0.133$, 95\% CI = [$-0.016$, $0.123$]) relative to control, respectively. These effects {for the accuracy nudge} are driven by decreases in false sharing of 2.3~pp (s.e. = $1.0$, $Z = -2.31$, $p = 0.021$, 95\% CI = [$-4.2$, $-0.35$])
, equivalent to {a} 4.9\% 
reduction in false sharing relative to control. Effects on true sharing are not distinguishable from zero for either treatment {(accuracy nudge estimate = $0.8$~pp, s.e. = $0.9$, $Z = 0.87$, $p = 0.387$, 95\% CI = [$-1.04$, $2.67$]; Facebook tips estimate = $0.5$~pp, s.e. = $1.0$, $Z = 0.47$, $p = 0.635$, 95\% CI = [$-1.54$, $2.53$])}. 

\begin{table}[H]
\centering
\resizebox{\textwidth}{!}{ 

\begin{tabular}[t]{lccccccc}
 & \textbf{Sharing} &  & \textbf{False} &  &  & \textbf{True} & \\
 & \textbf{Discernment} & Any sharing & Messenger & Timeline & Any sharing & Messenger & Timeline\\\cmidrule(lr){2-2} \cmidrule(lr){3-5} \cmidrule(lr){6-8} \multicolumn{4}{l}{\textbf{Headline treatment effects}} \rule{0pt}{1.2\normalbaselineskip}\\
\hspace{1em}Fact check & \num{-0.031} & \num{-0.006} & \num{-0.008} & \num{-0.003} & \num{0.001} & \num{0.001} & \num{-0.005}\\
 & (\num{0.036}) & (\num{0.011}) & (\num{0.011}) & (\num{0.011}) & (\num{0.011}) & (\num{0.011}) & (\num{0.011})\\
 & $p =$ \num{0.395} & $p =$ \num{0.608} & $p =$ \num{0.483} & $p =$ \num{0.785} & $p =$ \num{0.947} & $p =$ \num{0.915} & $p =$ \num{0.655}\\
 & {}[\num{-0.101}, \num{0.040}] & {}[\num{-0.028}, \num{0.016}] & {}[\num{-0.029}, \num{0.014}] & {}[\num{-0.024}, \num{0.018}] & {}[\num{-0.020}, \num{0.021}] & {}[\num{-0.021}, \num{0.023}] & {}[\num{-0.026}, \num{0.016}]\\
\hspace{1em}Related articles & \num{-0.052} & \num{0.008} & \num{0.002} & \num{0.011} & \num{-0.010} & \num{-0.006} & \num{-0.009}\\
 & (\num{0.035}) & (\num{0.011}) & (\num{0.011}) & (\num{0.011}) & (\num{0.011}) & (\num{0.011}) & (\num{0.011})\\
 & $p =$ \num{0.140} & $p =$ \num{0.477} & $p =$ \num{0.855} & $p =$ \num{0.279} & $p =$ \num{0.366} & $p =$ \num{0.568} & $p =$ \num{0.392}\\
 & {}[\num{-0.122}, \num{0.017}] & {}[\num{-0.014}, \num{0.030}] & {}[\num{-0.020}, \num{0.024}] & {}[\num{-0.009}, \num{0.032}] & {}[\num{-0.031}, \num{0.011}] & {}[\num{-0.028}, \num{0.015}] & {}[\num{-0.031}, \num{0.012}]\\\multicolumn{4}{l}{\textbf{Respondent treatment effects}} \rule{0pt}{1.2\normalbaselineskip}\\
\hspace{1em}Accuracy & \num{0.066} & \num{-0.023} & \num{-0.018} & \num{-0.025} & \num{0.008} & \num{0.011} & \num{0.002}\\
 & (\num{0.032}) & (\num{0.010}) & (\num{0.010}) & (\num{0.009}) & (\num{0.009}) & (\num{0.010}) & (\num{0.010})\\
 & $p =$ \num{0.038} & $p =$ \num{0.021} & $p =$ \num{0.060} & $p =$ \num{0.006} & $p =$ \num{0.387} & $p =$ \num{0.254} & $p =$ \num{0.874}\\
 & {}[\num{0.004}, \num{0.129}] & {}[\num{-0.042}, \num{-0.003}] & {}[\num{-0.037}, \num{0.001}] & {}[\num{-0.043}, \num{-0.007}] & {}[\num{-0.010}, \num{0.027}] & {}[\num{-0.008}, \num{0.030}] & {}[\num{-0.017}, \num{0.021}]\\
\hspace{1em}Facebook tips & \num{0.054} & \num{-0.020} & \num{-0.020} & \num{-0.016} & \num{0.005} & \num{0.010} & \num{0.002}\\
 & (\num{0.036}) & (\num{0.011}) & (\num{0.011}) & (\num{0.010}) & (\num{0.010}) & (\num{0.011}) & (\num{0.011})\\
 & $p =$ \num{0.133} & $p =$ \num{0.069} & $p =$ \num{0.069} & $p =$ \num{0.123} & $p =$ \num{0.635} & $p =$ \num{0.343} & $p =$ \num{0.819}\\
 & {}[\num{-0.016}, \num{0.123}] & {}[\num{-0.041}, \num{0.002}] & {}[\num{-0.041}, \num{0.002}] & {}[\num{-0.036}, \num{0.004}] & {}[\num{-0.015}, \num{0.025}] & {}[\num{-0.011}, \num{0.031}] & {}[\num{-0.019}, \num{0.024}]\\
\hspace{1em}Learned Targeted Policy & \num{0.069} & \num{-0.020} & \num{-0.021} & \num{-0.025} & \num{0.006} & \num{0.013} & \num{-0.001}\\
 \hspace{1.5em}(maximizing sharing discernment) & (\num{0.032}) & (\num{0.010}) & (\num{0.010}) & (\num{0.009}) & (\num{0.009}) & (\num{0.010}) & (\num{0.010})\\
 & $p =$ \num{0.030} & $p =$ \num{0.037} & $p =$ \num{0.031} & $p =$ \num{0.007} & $p =$ \num{0.509} & $p =$ \num{0.190} & $p =$ \num{0.915}\\
 & {}[\num{0.007}, \num{0.131}] & {}[\num{-0.040}, \num{-0.001}] & {}[\num{-0.039}, \num{-0.002}] & {}[\num{-0.043}, \num{-0.007}] & {}[\num{-0.012}, \num{0.024}] & {}[\num{-0.006}, \num{0.031}] & {}[\num{-0.020}, \num{0.018}]\\
\hspace{1em}Restricted Targeted Policy & \num{0.095} & \num{-0.033} & \num{-0.031} & \num{-0.035} & \num{0.005} & \num{0.009} & \num{-0.001}\\
 \hspace{1.5em}(minimizing any false sharing) & (\num{0.033}) & (\num{0.010}) & (\num{0.010}) & (\num{0.009}) & (\num{0.010}) & (\num{0.010}) & (\num{0.010})\\
 & $p =$ \num{0.004} & $p =$ \num{0.001} & $p =$ \num{0.002} & $p =$ \num{<0.001} & $p =$ \num{0.605} & $p =$ \num{0.339} & $p =$ \num{0.911}\\
 & {}[\num{0.030}, \num{0.160}] & {}[\num{-0.052}, \num{-0.013}] & {}[\num{-0.050}, \num{-0.011}] & {}[\num{-0.053}, \num{-0.016}] & {}[\num{-0.014}, \num{0.024}] & {}[\num{-0.010}, \num{0.029}] & {}[\num{-0.020}, \num{0.018}]\\\hline
\hspace{1em}Control mean & \num{-0.403} & \num{0.466} & \num{0.419} & \num{0.387} & \num{0.679} & \num{0.586} & \num{0.615}\\
 & (\num{0.027}) & (\num{0.008}) & (\num{0.008}) & (\num{0.008}) & (\num{0.008}) & (\num{0.008}) & (\num{0.008})\\
\end{tabular}

\vspace{1ex}

}
\caption{\textbf{Control response and treatment effect estimates.} The sample is users in the evaluation stage, $n = 10,531$. Columns denote response measures, described in the note to \autoref{tab:heterogeneity_control}. 
The last row represents estimated mean response under the control condition; all other rows are estimated treatment effects in contrast with the control condition. Estimates are produced from an augmented inverse probability weighted estimator, as described in \autoref{section:estimation}. 
{Standard errors are presented beneath estimates in parentheses; for treatment effects, two-sided p-values and confidence intervals follow.}
Alternative specifications are presented in \autoref{tab:alternative_results}. 
} 
\label{tab:main_results}

\end{table}

\paragraph{Heterogeneity in best policy}\label{section:het_policy}
{While we observe directional similarities in the effects of the accuracy nudge and Facebook tips,} we also observe differences in how users respond to these treatments. The Learned Targeted Policy reported in \autoref{tab:main_results} shows modest improvements over control{, but does not meaningfully improve over uniformly assigning the accuracy nudge or Facebook tips on sharing discernment and false sharing intentions}. The Restricted Targeted Policy, however, illustrates the benefits of personalization: which treatment is most effective varies across users. 

The Restricted Targeted Policy achieves a decrease in false intentions sharing relative to control of $-3.3$~pp (s.e. = $1.0$, $Z = -3.23$, $p = 0.001$, 95\% CI = [$-5.23$, $-1.28$]) (see \autoref{tab:main_results}), the objective it was trained to optimize using data from the learning stage. This 7\% reduction is an improvement as compared to either the accuracy nudge (estimate = $-1.0$~pp, s.e. = $0.4$, $Z = -2.53$, $p = 0.011$, 95\% CI = [$-1.75$, $-0.22$]) 
assigned uniformly, {or the Learned Targeted Policy (estimate = $-1.2$~pp, s.e. = $0.4$, $Z = -3$, $p = 0.003$, 95\% CI = [$-2.01$, $-0.42$])}. 

{The Restricted Targeted Policy also improves the discernment outcome relative to the accuracy nudge assigned uniformly (estimate = $0.029$, s.e. = $0.013$, $Z = 2.145$, $p = 0.032$, 95\% CI = [$0.002$, $0.055$]); 
improvements relative to the Learned Targeted Policy are not statistically distinguishable from zero
(estimate = $0.026$, s.e. = $0.013$, $Z = 1.932$, $p = 0.053$, 95\% CI = [$0$, $0.052$]).  
It might seem surprising that even though the Restricted Targeted Policy was trained to optimize reduction in false sharing, we are able to achieve improvements over the accuracy nudge on the discernment outcome under this policy, while we were not with the Learned Targeted Policy. This finding can be understood in light of the fact that attempting to estimate a policy with many potential treatments may result in overfitting or high-variance estimates of the best policy, and so we may benefit from reducing the number of possible assignments from the Learned Targeted Policy from four to two under the Restricted Targeted Policy. Further, treatment effect heterogeneity is primarily driven by heterogeneity in false sharing, and so by targeting assignment only according to the false sharing measure, we may be using a stronger signal.}

\autoref{tab:heterogeneity_best} presents results by assignment group within the Restricted Targeted Policy. We see the policy has appropriately assigned participants to the respective respondent-level conditions: on average, participants assigned to receive the accuracy nudge under the policy intend to share false information at 4.9\% lower rates under the accuracy nudge as compared to Facebook tips{, although this difference is not statistically distinguishable from zero} (estimate = $-2.1$~pp, s.e. = $1.2$, $Z = -1.80$, $p = 0.071$, 95\% CI = [$-4.41$, $0.18$]); the reverse is true for participants assigned to Facebook tips, who see an increase of 13.3\% (estimate = $6.5$~pp, s.e. = $2.3$, $Z = 2.81$, $p = 0.005$, 95\% CI = [$1.97$, $11.06$]). Because the Restricted Targeted Policy is optimized to minimize false sharing, we see smaller relative differences in true sharing. 

\begin{table}[H]
\small
   \centering
\resizebox{\textwidth}{!}{ 

\begin{tabular}[t]{lccccccc}
 & \textbf{Sharing} &  & \textbf{False} &  &  & \textbf{True} & \\
 & \textbf{Discernment} & Any sharing & Messenger & Timeline & Any sharing & Messenger & Timeline\\\cmidrule(lr){2-2} \cmidrule(lr){3-5} \cmidrule(lr){6-8} \multicolumn{4}{l}{\textbf{Optimal assignment == Accuracy nudge (n = 8,309)}} \rule{0pt}{1.2\normalbaselineskip}\\
\hspace{1em}Accuracy & \num{-0.237} & \num{0.413} & \num{0.368} & \num{0.338} & \num{0.682} & \num{0.584} & \num{0.617}\\
 & (\num{0.024}) & (\num{0.008}) & (\num{0.007}) & (\num{0.007}) & (\num{0.007}) & (\num{0.007}) & (\num{0.007})\\
\hspace{1em}Facebook Tips & \num{-0.303} & \num{0.434} & \num{0.382} & \num{0.365} & \num{0.684} & \num{0.590} & \num{0.619}\\
 & (\num{0.029}) & (\num{0.009}) & (\num{0.009}) & (\num{0.009}) & (\num{0.009}) & (\num{0.009}) & (\num{0.009})\\
\cmidrule(lr){2-8}
\hspace{1em}Difference & \num{0.066} & \num{-0.021} & \num{-0.015} & \num{-0.027} & \num{-0.002} & \num{-0.006} & \num{-0.002}\\
 & (\num{0.038}) & (\num{0.012}) & (\num{0.012}) & (\num{0.011}) & (\num{0.011}) & (\num{0.012}) & (\num{0.012})\\
 & $p =$ \num{0.081} & $p =$ \num{0.071} & $p =$ \num{0.210} & $p =$ \num{0.016} & $p =$ \num{0.846} & $p =$ \num{0.604} & $p =$ \num{0.865}\\
 & {}[\num{-0.008}, \num{0.140}] & {}[\num{-0.044}, \num{0.002}] & {}[\num{-0.037}, \num{0.008}] & {}[\num{-0.049}, \num{-0.005}] & {}[\num{-0.024}, \num{0.020}] & {}[\num{-0.029}, \num{0.017}] & {}[\num{-0.025}, \num{0.021}]\\\multicolumn{4}{l}{\textbf{Optimal assignment == Facebook tips (n = 2,222)}} \rule{0pt}{1.2\normalbaselineskip}\\
\hspace{1em} Accuracy & \num{-0.710} & \num{0.555} & \num{0.523} & \num{0.451} & \num{0.705} & \num{0.650} & \num{0.614}\\
 & (\num{0.043}) & (\num{0.014}) & (\num{0.014}) & (\num{0.014}) & (\num{0.013}) & (\num{0.013}) & (\num{0.013})\\
\hspace{1em} Facebook Tips & \num{-0.524} & \num{0.490} & \num{0.461} & \num{0.396} & \num{0.682} & \num{0.623} & \num{0.611}\\
 & (\num{0.062}) & (\num{0.019}) & (\num{0.019}) & (\num{0.017}) & (\num{0.016}) & (\num{0.017}) & (\num{0.016})\\\cmidrule(lr){2-8}
\hspace{1em} Difference & \num{-0.186} & \num{0.065} & \num{0.062} & \num{0.055} & \num{0.024} & \num{0.027} & \num{0.003}\\
 & (\num{0.075}) & (\num{0.023}) & (\num{0.024}) & (\num{0.022}) & (\num{0.020}) & (\num{0.021}) & (\num{0.021})\\
 & $p =$ \num{0.014} & $p =$ \num{0.005} & $p =$ \num{0.009} & $p =$ \num{0.012} & $p =$ \num{0.246} & $p =$ \num{0.198} & $p =$ \num{0.888}\\
 & {}[\num{-0.334}, \num{-0.038}] & {}[\num{0.020}, \num{0.111}] & {}[\num{0.016}, \num{0.108}] & {}[\num{0.012}, \num{0.099}] & {}[\num{-0.016}, \num{0.063}] & {}[\num{-0.014}, \num{0.069}] & {}[\num{-0.038}, \num{0.044}]\\
\end{tabular}

   }
   \caption{\textbf{Response under counterfactual uniform respondent treatment conditions, by {Restricted Targeted Policy} assignment.} The sample is users in the evaluation stage, $n = 10,531$. Estimates are of mean response under the two respondent-level treatments. 
   Columns denote response measures, described in the note to \autoref{tab:heterogeneity_control}. 
   Estimates are produced from an augmented inverse probability weighted estimator, as described in \autoref{section:estimation}, within specified subgroups. 
   {Standard errors are presented beneath estimates in parentheses; for differences, two-sided p-values and confidence intervals follow.}
   }
   \label{tab:heterogeneity_best}
   
\end{table}

Previous research has asked 
whether accuracy nudges and Facebook tips impact behavior through the mechanism of increasing attention to accuracy, as suggested for the accuracy nudge by ref. \citenum{pennycook_epstein_mosleh_arechar_eckles_rand_2019}, 
or rather whether Facebook tips improve ability to evaluate stimuli, as proposed by ref. \citenum{guessetal2020digital}. {The heterogeneity we find in treatment effects between the two groups (estimate = $8.6$, s.e. = $2.6$, $Z = 3.23$, $p = 0.001$, 95\% CI = [$-1.37$, $-0.35$]) suggests that different types of people respond differently to each treatment.} 

\begin{figure}[H] 
   \centering
   \includegraphics[width=\textwidth]{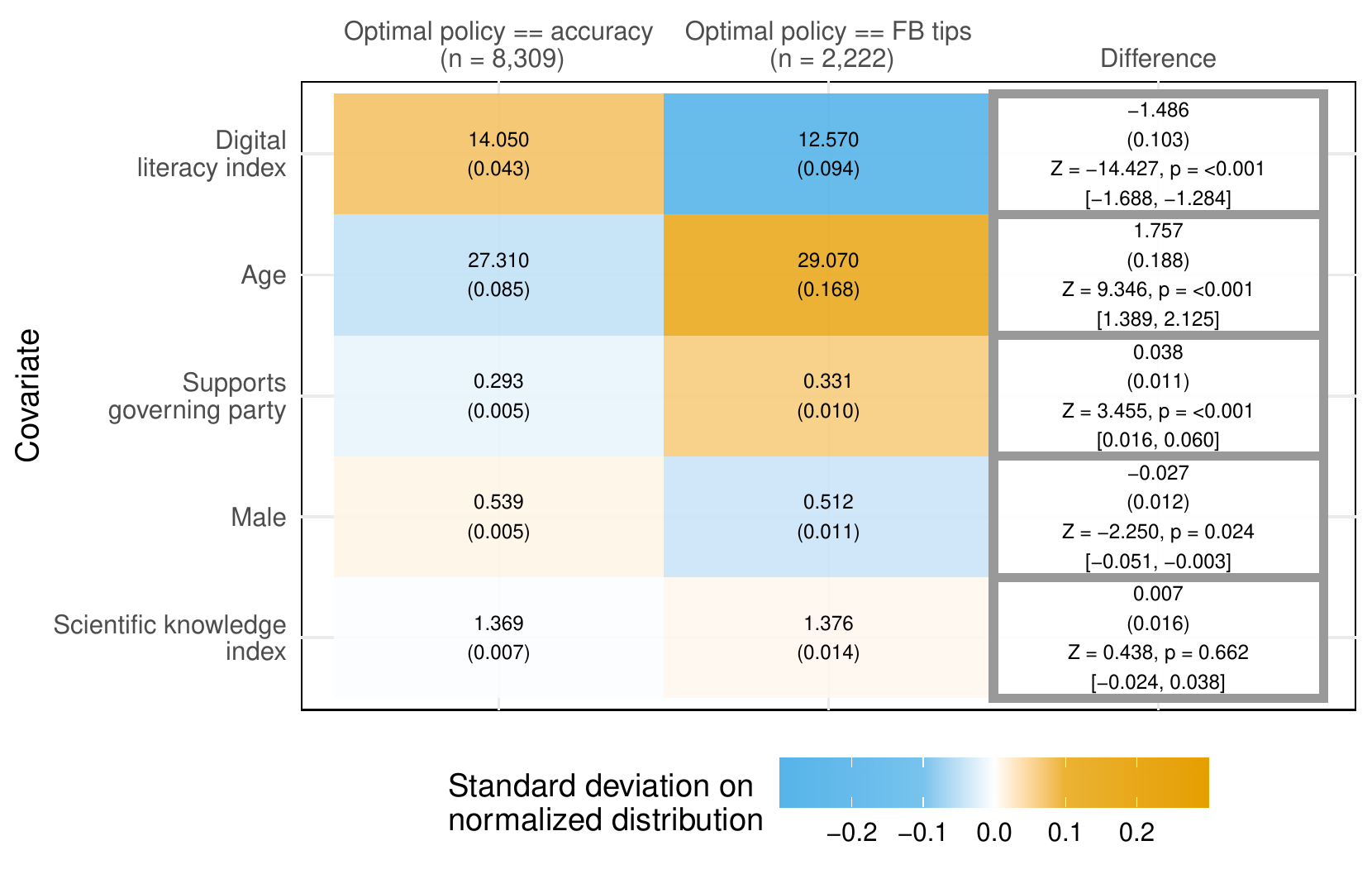} 
   \caption{\textbf{Selected covariate means by Restricted Targeted Policy assignment.} The sample is users in the evaluation stage, $n = 10,531$. Covariates are ordered by size of standardized deviation between the two groups. 
   }
   \label{fig:heterogeneity_covariates}

\end{figure}

In \autoref{fig:heterogeneity_covariates}, we report differences in selected covariates across the groups assigned to each treatment in the Restricted Targeted Policy. 
The 78.9\% of participants assigned to the accuracy nudge are, on average, more digitally literate and slightly younger; they are also less likely to support the governing party and more likely to be male.

\paragraph{Sharing channel}
To further investigate how the accuracy nudge 
operates, we consider the secondary dimension of our response measurement: sharing channel. The accuracy nudge 
effectively reduces false sharing intentions on Timeline (estimate = $-2.5$~pp, s.e. = $0.9$, $Z = -2.76$, $p = 0.006$, 95\% CI = [$-4.34$, $-0.74$]
)
{and directionally} reduces false sharing intentions on Messenger (estimate = $-1.8$~pp, s.e. = $1.0$, $Z = -1.88$, $p = 0.060$, 95\% CI = [$-3.68$, $0.08$]
)
relative to control{, although this latter difference is not statistically distinguishable from zero}. 

\section{Discussion}\label{sec:discussion}

This study provides evidence from two of the largest Facebook populations in sub-Saharan Africa that online interventions delivered via Facebook Messenger are effective at improving sharing discernment.  
This study brings comparative data to the global problem of health misinformation, which to date draws primarily on empirical evidence from samples in the US, Canada, and Europe.

{With respect to our discernment outcome, we estimate (precise) null effects of the headline-level interventions relative to the control. 
We consider the tested headline-level treatments in the context of the literature: Only a handful of scholars have previously examined Facebook's related articles policy \citep{bode2015related}, hence more research is needed to assess whether our null estimate is context-specific or whether it replicates across different settings. In contrast, numerous experimental studies find fact checks to be effective at reducing users' belief in false stories \citep{nyhan2010corrections,clayton2020real,brashier2021timing,porter2021global}, but few focus on whether users intend to share the information. 
Our null results on sharing lend further support to the notion that what users believe is distinct from, although related to, what they share \citep{epstein2023social}.}

{We do, however, see that on average, individuals share more discerningly when given the accuracy nudge.  
This study contributes to others that have found positive effects of accuracy prompts, including among quota-matched samples in 16 countries \citep{arechar2022understanding} and in a meta-analysis of 20 studies with a total sample size over 20,000 \citep{pennycook2022accuracy}.}
{For a subset of the sample, we find that Facebook tips is more effective than the accuracy nudge. }
{Facebook tips have also been shown to reduce belief in false headlines in the US and India \citep{guessetal2020digital}, indicating that both treatments may be scalable solutions for the global misinformation challenge. 
} 

{A fruitful avenue for future work would be to explore differences in treatment effects on sharing discernment across Messenger and Timeline to better understand mechanisms. For example, if the accuracy nudge reminds users to pay attention to the accuracy of posts, it may also emphasize to users that their peers value accuracy. As a result, this treatment may make users particularly wary of sharing false posts publicly on their Timeline given reputational concerns and fear of repercussions from peers for sharing misinformation \citep{altay2022so}.}

This study has limitations.  First, we aimed to identify interventions that are effective among social media users in Kenya and Nigeria. We were limited, however, to engaging with those who clicked on our Facebook advertisements. Recruiting actual social media users on the platform has advantages in validity relative to convenience samples, laboratory experiments, and opt-in survey panels. We cannot say, however, how users who decided to participate in our study differ on unobservables from general populations in these countries. 

Second, interacting with study participants and delivering interventions in a survey experiment cannot perfectly capture how users would react to interventions delivered on the platform. Although still artificial, delivering the survey and interventions through Messenger conversations provides greater realism than interventions delivered on third party survey platforms. The nature of our experiment means participants were aware they were part of a study (rather than an on-platform field experiment,  where consent may be waived by IRB or implicitly provided when users agree to the terms and conditions). It is possible participants' responses were driven by experimenter demand effects. 

To address experimenter demand effects, we embedded treatments in a longer survey block about general social media usage. If users' posttreatment responses were based on perceptions of what researchers want, we might expect high digital literacy users to be the most savvy to survey objectives and treatment effects to be largest for this group. Instead, we observe the reverse. 

Finally, misinformation studies that focus on sharing outcomes are constrained by ethical considerations of contributing to the ecosystem of misinformation when looking at real online sharing behavior. Instead this study, like others, used measures of sharing intentions.  While stated intentions correlate with online sharing behavior \citep{mosleh2020self}, measuring intentions rather than real behavior remains a limitation of scholarship in this area.  

Acknowledging these limitations,
{these findings have implications for fighting misinformation generally, and health misinformation specifically. 
}
Low-cost and scalable interventions like accuracy nudges can be effective in diverse contexts. 
This study provides evidence that such interventions are more effective than others tested by researchers and used by platforms. 
Platforms may be more likely to deliver such interventions knowing they help reduce sharing of misinformation without hindering sharing of true information. 

\section{Methods}\label{section:methods}

\subsection{Data and recruitment}\label{section:data}
\if0\blind{%
{This study was approved by the Stanford IRB, protocol: 57430. This study was also covered by local IRBs in Kenya and Nigeria under the Busara Center for Behavioral Economics IRB approvals for psychological and behavioral studies.}
}\fi
\if1\blind{%
[IRB information blinded in this version.]  }\fi
\if0\blind{%
The design for this study is preregistered on the Open Science Framework registry: \url{https://osf.io/ny2xc}.}\fi
\if1\blind{%
The design for this study is preregistered on the Open Science Framework registry.}\fi

Our sample was recruited from Facebook users in Kenya and Nigeria, two of Facebook's top-three largest user bases in sub-Saharan Africa \citep{africa2016top}, with a combined user base of 30--35 million users aged 18 years and older %
(as reported on the audience insights tool on Facebook's advertising platform). %
We used targeted Facebook advertisements to improve balance on age and gender. After users clicked on our advertisements offering airtime for taking a survey, 
they consented to participate in this study, and started a conversation with our page's Messenger chatbot. 

Users who completed the survey received compensation in the form of mobile phone airtime (equivalent to about \$0.50) sent to their phone. {All participants were told at the outset how much the incentive was for their participation.} {This survey incentive was selected in consultation with local research firms and adheres to minimum wage regulations in each country. In Nigeria, minimum wage is 30,000 NGN (\$65 USD) per month, equivalent to \$0.41 USD per hour (\url{https://mywage.ng/salary/minimum-wage-1}). In Kenya, minimum hourly wage ranges from 76 KSH (\$0.59 USD) to 309 KSH (\$2.40 USD) per hour based on skill level and location. Hence, offering \$0.50 for 20 minutes seemed reasonable without being excessive. Offering excessive incentives can lead to respondents attempting to take the survey multiple times, sometimes with great tenacity (e.g., creating new Facebook accounts).} 

{We observed posttreatment attrition of 5.6\% in the learning stage and 8.7\% in the evaluation stage. We discuss estimation techniques to account for attrition in \autoref{section:estimation}.} {{18,733 total respondents engaged with our chatbot in the learning stage, and 27,758 respondents in the evaluation stage. Of these, 12,354 in the learning stage and 20,174 in the evaluation stage were eligible and consented to participate in the survey; 5,067 users in the learning stage and 11,567 in the evaluation stage remained in the survey long enough to be assigned treatment. We removed an additional 24 users from the learning stage and 33 in the evaluation stage for implementation errors in assigning treatment due to server failures. We assume that attrition to this point is orthogonal to (unrealized) treatment assignment. We were able to collect posttreatment responses for 4,761 users in the learning stage and 10,531 in the evaluation stage; these 15,292 constitute the total sample used for analysis here. Power calculations for both the learning and evaluation samples are reported in our online preregistration. } }


\paragraph{Stimuli}\label{section:measures}
Each participant saw four {pretreatment and four} posttreatment stimuli, two true and two false in a random order. 
{At the user level, stimuli were randomly selected from our database so that each individual saw eight unique stimuli. }
The stimuli include true information, sourced from the WHO, the Nigeria Center for Disease Control, the National Emergency Response Committee in Kenya, and each country's Ministry of Health. The false posts were sourced from AFP, Poynter, and AfricaCheck websites' lists of online misinformation; {these resources are now available in a unified database at \url{https://www.poynter.org/ifcn-covid-19-misinformation/}.} The misinformation was fact checked in Kenya and Nigeria since the start of the pandemic. 


\paragraph{Treatments}\label{section:treatments}

We considered two types of treatments, both randomized at the user-level: headline-level interventions applied to stimuli, and respondent-level interventions targeted to the participants themselves. In the evaluation stage, we tested two of each type of intervention against control, along with a Learned Targeted Policy composed of four of the respondent-level treatments. 

\begin{figure}[H]
\centering
\includegraphics[width=\textwidth]{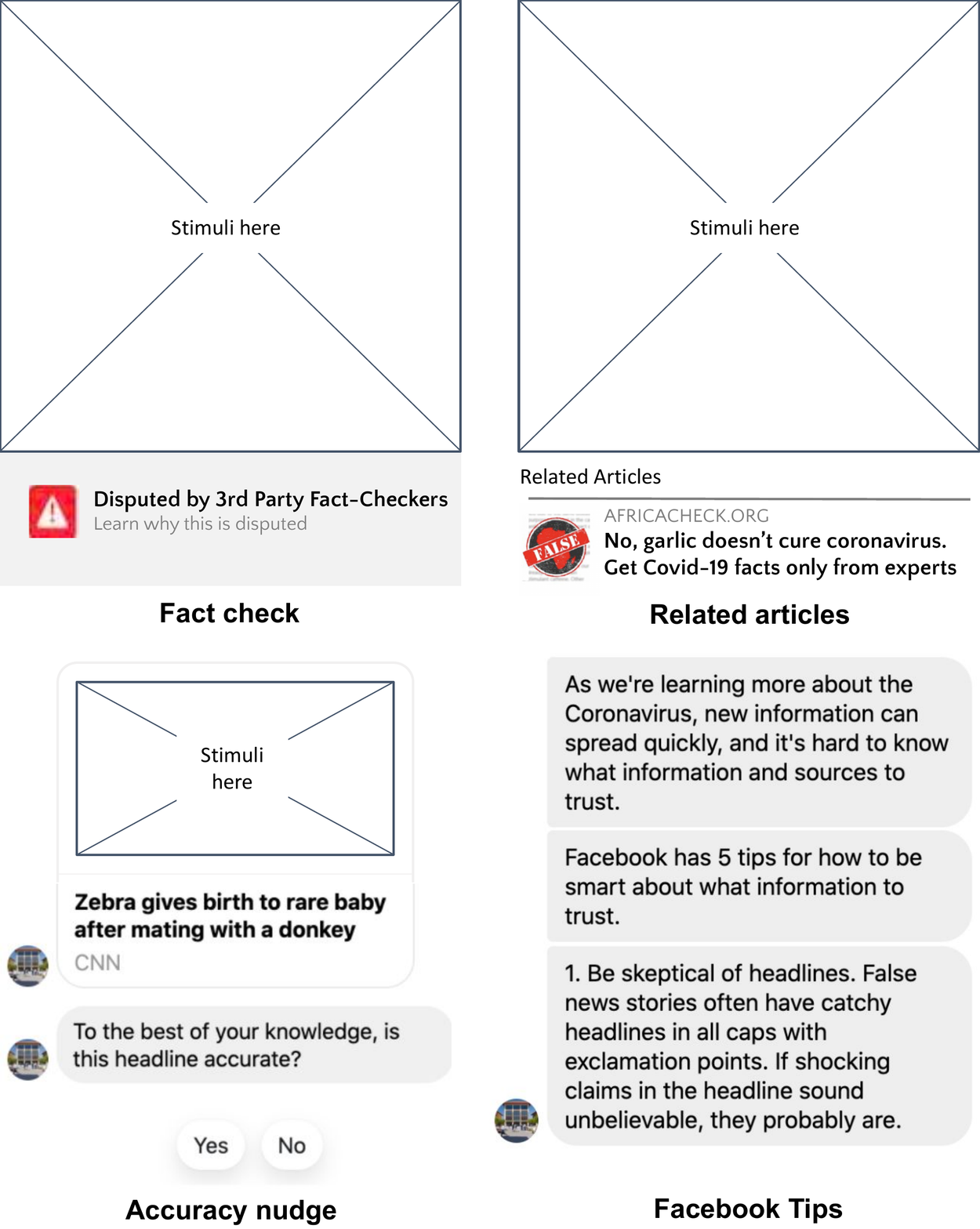}
\caption{\textbf{Headline- and respondent-level treatments tested in the evaluation stage.} The first row depicts headline-level treatments, the second row respondent-level treatments. All stimuli used are available at \url{http://bit.ly/facebook_stimuli_public}. Specific text in treatments may vary based on stimuli used.}
\label{fig:4treatments}

\end{figure}

The selected uniform treatments were the accuracy nudge and Facebook tips (respondent-level) and fact checks and related articles (headline-level). The accuracy nudge asked participants to tell us whether they thought a separate post, unrelated to COVID, was accurate \citep{pennycook2020fighting}. The Facebook tips treatment provided participants with ten tips from Facebook on how to be smart about what information to trust, including being skeptical of headlines, watching for unusual formatting, checking the evidence, and looking at other reports. The full text of Facebook's tips is presented in Supplementary~\autoref*{appendix:treatments}. The fact check treatment included a warning label on false stimuli, modeled on one used by Facebook for its third-party fact checking program. The related articles treatment was also modeled on a program tested by Facebook, which paired disputed articles with articles on the same topic from validated sources \citep{ghosh2017facebook}. Examples of each are presented in \autoref{fig:4treatments}. 

{In interpreting results of the headline-level treatments, we assume respondents thought—as one might on Facebook—that they were sharing the entire content (original post plus fact check). Some users likely wanted to share the post to share the fact check, rather than spread misinformation. The ambiguity involved in what users would actually be sharing reflects the ambiguity on these platforms. If we care most about reducing any sharing of false information, we would also want to limit sharing of false posts even when accompanied by fact checks/related articles. Indeed, Facebook reprimands sharing of fact-checked false posts. (\href{https://transparency.fb.com/enforcement/taking-action/penalties-for-sharing-fact-checked-content/}{https://transparency.fb.com/enforcement/taking-action/penalties-for-sharing-fact-checked-content/}) Mirroring the ambiguity of these interventions run on social media platforms, this intervention is helpful in understanding how such flags affect sharing rates generally, whatever the respondent's motivation, compared to other interventions.
}

\subsection{Empirical strategy}\label{section:estimation}


All estimates reported are calculated separately for the learning and evaluation data. The learning data are used for learning which treatment conditions have highest mean response, and estimating targeted treatment assignment policies, but we evaluate these targeted policies and our other main experimental effects separately on the evaluation data. 

For the estimates of average response under the different treatment conditions and average treatment effects reported from the evaluation data, we use a generalized augmented inverse probability weighted estimator \citep{robins1994estimation}. The estimator accounts for both weighting for unequal treatment assignment probabilities and adjusting for covariates. 

Each individual's scores under treatment condition $w$ for the augmented inverse probability weighted estimator are calculated as
\begin{equation}
\Gamma^{AIPW}_i(w) := \hat \mu_i (X_i; w) + \frac{\textbf{1}\{W_i = w\}}{e_i(X_i;w)}\left(Y_i - \hat\mu_i(X_i;w) \right),
\end{equation}
where observed response for individual $i$ is represented by $Y_i$, assigned treatment probabilities are represented by $e_i(X_i;w):=\Pr[W_i=w\mid X_i=x]$; and $\hat \mu_i (X_i; w)$ is a conditional means model, with covariates $X_i$ and categorical treatments $W_i \in \W$. We estimate the conditional means model using a random forest as implemented by the \texttt{grf} page in R statistical software, which ensures that the outcome for unit $i$ is not used in constructing the prediction for unit $i$  \citep{Tibshirani:2020aa}.

The estimator is a substitution estimator, so we can predict counterfactual response for each individuals under each treatment condition. We estimate average response under each condition by taking the averages of respective scores across individuals, estimating $\textrm{E}[Y_i(w)]$ as
\begin{equation}
Q_i^{AIPW}(w) := \frac{1}{N}\sum_{i = 1}^N \Gamma^{AIPW}_i(w). 
\end{equation}
For contrasts, we take average differences in scores $\Gamma^{AIPW}_i(w)-\Gamma^{AIPW}_i(w')$. For estimates of heterogeneity, we condition only on the relevant subset of the sample when averaging scores. Estimation of standard errors follows the implementation in ref. \citenum{Tibshirani:2020aa}.

We use a similar approach to estimation for learning data using the augmented inverse probability weighted estimator. However, we must also account for nonnormality of the estimator on the adaptively collected data. To do so, we use adaptive weights, described in ref. \citenum{zhan2021off}. We estimate $\textrm{E}[Y_i(w)]$ from the learning stage, using weights $h_i(w)$, as
\begin{equation}
Q_i^{h}(w) := \frac{ \frac{1}{N}\sum_{i = 1}^N h_i(w)\Gamma^{AIPW}_i(w)}{\sum_{i = 1}^N h_i(w)} . 
\label{eq:scores_learning}
\end{equation}
The weights are produced as the contextual stabilized variance weights. For the learning data, we also predict $\hat \mu_i (X_i; w)$ for each observation using only historical data. 

 Covariates used for adjustment in both the learning and evaluation stages are described in further detail in \autoref*{tab:cov_long}. 
{We do not account for attrition in the learning stage, as our policy learning in the experiment was conducted only on the fully collected learning stage data. For evaluation stage estimates, we account for attrition by weighting completed survey responses to the sample of evaluation stage users assigned treatment and estimate censoring probabilities by conditioning on all pretreatment covariates.}

Data collection and analysis were not performed blind to the conditions of the experiments.


\section{Data availability statement}
\if0\blind{%
The data that support the findings of this study are available at \url{https://github.com/gsbDBI/infodemic-replication}. 
}\fi
\if1\blind{%
The data that support the findings of this study are available at \url{https://anonymous.4open.science/r/infodemic-replication-5333}.
}\fi

\section{Code availability statement}
\if0\blind{%
The analysis code that generate the figures, tables, and results presented in this study are available at \url{https://github.com/gsbDBI/infodemic-replication}.
}\fi
\if1\blind{%
The analysis code that generate the figures, tables, and results presented in this study are available at \url{https://anonymous.4open.science/r/infodemic-replication-5333}.
 }\fi

\section{Acknowledgements}
\if0\blind{%
We received advertising credits for this study from Facebook Health and funding from the Golub Capital Social Impact Lab and Office of Naval Research grant N00014-19-1-246 (S.A.). The funders had no role in study design, data collection and analysis, decision to publish or preparation of the manuscript. For exceptional research assistance, we thank Zelin (James) Li, Ricardo Ruiz, Undral Byambadalai, and Haotian (Tony) Zong. We thank Justine Davis, Shelby Grossman, Laura Jakli, Edward Jee, Tanu Kumar, Emil Palikot, and Alex Siegel for feedback and comments, as well as the participants of the seminar series of the Development Innovation Lab at the Becker Friedman Institute. 
We thank James Kiselik for editorial assistance. }\fi
\if1\blind{%
[blinded] }\fi

\section{Author contributions}
\if0\blind{%
M.O.W., L.R. and S.A. designed the research and wrote the paper. M.O.W. and L.R. performed the experimental studies. M.O.W. analyzed the data, with input from S.A. and L.R.. }\fi
\if1\blind{%
[blinded] }\fi

\section{Competing interests}
The authors declare no competing interests.


\bibliography{main.bib}

\clearpage



\appendix
\phantomsection\label{SI}
\renewcommand\thefigure{\arabic{figure}} 
\renewcommand\thetable{\arabic{table}} 
\renewcommand{\figurename}{Supplementary Figure}
\renewcommand{\tablename}{Supplementary Table}
\renewcommand{\figureautorefname}{Supplementary Figure}
\renewcommand{\tableautorefname}{Supplementary Table}

 \renewcommand{\thesection}{S\arabic{section}}
 \pagenumbering{arabic}
\renewcommand*{\thepage}{SI.\arabic{page}}

\setcounter{figure}{0} 
\setcounter{table}{0} 
\appendixpage


\section{Supplementary Methods}




\subsection{Sample Characteristics}\label{appendix:demographics}

\begin{table}[H]
\begin{center}
  \subfloat[\label{tab:abKY}Kenya]{\resizebox{\textwidth}{!}{ 

\begin{tabular}[t]{rcccc|ccccc}
 & Learning Mean & Learning SE & Evaluation Mean & Evaluation SE & Overall Facebook Mean & Facebook SE & Afrobarometer Mean & Afrobarometer SE & Difference\\
\cmidrule(lr){2-2}  \cmidrule(lr){3-3}  \cmidrule(lr){4-4} \cmidrule(lr){5-5} \cmidrule(lr){6-6} \cmidrule(lr){7-7} \cmidrule(lr){8-8} \cmidrule(lr){9-9} \cmidrule(lr){10-10}
Age & 30.81 & 0.34 & 28.87 & 0.09 & 29.43 & 0.09 & 36.16 & 0.34 & -6.73\\
Has cash income & 0.42 & 0.01 & 0.36 & 0.01 & 0.38 & 0.01 & 0.48 & 0.01 & -0.10\\
Education level & 7.16 & 0.05 & 7.18 & 0.02 & 7.18 & 0.02 & 4.85 & 0.05 & 2.33\\
Index of household possessions & 3.80 & 0.04 & 3.82 & 0.02 & 3.82 & 0.02 & 2.93 & 0.04 & 0.88\\
Male & 0.48 & 0.01 & 0.57 & 0.01 & 0.54 & 0.01 & 0.50 & 0.01 & 0.04\\
Supports governing party & 0.13 & 0.01 & 0.13 & 0.00 & 0.13 & 0.00 & 0.30 & 0.01 & -0.17\\
Christian & 0.91 & 0.01 & 0.90 & 0.00 & 0.91 & 0.00 & 0.76 & 0.01 & 0.14\\
Muslim & 0.04 & 0.01 & 0.04 & 0.00 & 0.04 & 0.00 & 0.09 & 0.01 & -0.06\\
Urban & 0.44 & 0.01 & 0.40 & 0.01 & 0.41 & 0.01 & 0.36 & 0.01 & 0.05\\
n & 2,180 &   & 5,318 &   & 7,498 &  & 1,599 &   &  
\end{tabular}

\vspace{1ex}
}}%
  \qquad
  \subfloat[\label{tab:abNG}Nigeria]{\resizebox{\textwidth}{!}{ 

\begin{tabular}[t]{rcccc|ccccc}
 & Learning Mean & Learning SE & Evaluation Mean & Evaluation SE & Overall Facebook Mean & Facebook SE & Afrobarometer Mean & Afrobarometer SE & Difference\\
\cmidrule(lr){2-2}  \cmidrule(lr){3-3}  \cmidrule(lr){4-4} \cmidrule(lr){5-5} \cmidrule(lr){6-6} \cmidrule(lr){7-7} \cmidrule(lr){8-8} \cmidrule(lr){9-9} \cmidrule(lr){10-10}
Age & 26.53 & 0.31 & 26.48 & 0.09 & 26.49 & 0.09 & 32.66 & 0.31 & -6.16\\
Has cash income & 0.41 & 0.01 & 0.39 & 0.01 & 0.40 & 0.01 & 0.49 & 0.01 & -0.09\\
Education level & 7.20 & 0.05 & 7.39 & 0.02 & 7.33 & 0.02 & 5.52 & 0.05 & 1.82\\
Index of household possessions & 4.58 & 0.04 & 4.47 & 0.02 & 4.50 & 0.02 & 4.00 & 0.04 & 0.50\\
Male & 0.54 & 0.01 & 0.50 & 0.01 & 0.51 & 0.01 & 0.50 & 0.01 & 0.01\\
Supports governing party & 0.22 & 0.01 & 0.17 & 0.00 & 0.18 & 0.00 & 0.26 & 0.01 & -0.07\\
Christian & 0.63 & 0.01 & 0.69 & 0.01 & 0.67 & 0.01 & 0.55 & 0.01 & 0.12\\
Muslim & 0.33 & 0.01 & 0.27 & 0.01 & 0.29 & 0.01 & 0.42 & 0.01 & -0.13\\
Urban & 0.59 & 0.01 & 0.61 & 0.01 & 0.60 & 0.01 & 0.44 & 0.01 & 0.17\\
N & 2,581 &   & 5,213 &   & 7,794 &  & 1,600 &   &  
\end{tabular}

\vspace{1ex}
}}
\end{center}
\caption{\textbf{Comparing Facebook and Afrobarometer samples.} 
{The learning and evaluation samples are recruited on Facebook, and means are reported separately and overall. Differences are between the overall Facebook sample and the Afrobarometer sample.} The Facebook sample household asset index is re-coded to exclude a question about bike ownership, to match the household asset index in the Afrobarometer data. Analysis in the main paper includes this additional question in the index. Additionally, the Facebook sample governing party support variable is coded to only include affiliation with the governing party here, again to match the Afrobarometer data, which only asks about prospective voting. In the main analysis, the governing party support variable is coded as 1 if the respondent either responds that they feel affiliation with the governing party, or if they voted for a candidate from that party in the previous election.} 
  \refstepcounter{SItable}\label{tab:ab}
\end{table}

\begin{table}[H] 
\begin{center}
  \subfloat[\label{tab:sharesT}True shares]
{
\begin{tabular}[t]{rccccc}
Shares & Batch 1 & Batch 2 & Batch 3 & Batch 4 & Batch 5\\
 \cmidrule(lr){2-2}  \cmidrule(lr){3-3}  \cmidrule(lr){4-4} \cmidrule(lr){5-5} \cmidrule(lr){6-6} 
0 & 0.16 & 0.19 & 0.18 & 0.17 & 0.15\\
1 & 0.08 & 0.07 & 0.08 & 0.06 & 0.09\\
2 & 0.18 & 0.18 & 0.16 & 0.15 & 0.18\\
3 & 0.11 & 0.10 & 0.10 & 0.09 & 0.10\\
4 & 0.47 & 0.46 & 0.47 & 0.52 & 0.48\\
N & 2,507 & 827 & 804 & 905 & 11,534
\end{tabular}

\vspace{1ex}
}%
  \qquad\normalsize
  \subfloat[\label{tab:sharesF}False shares]{

\begin{tabular}[t]{rccccc}
Shares & Batch 1 & Batch 2 & Batch 3 & Batch 4 & Batch 5\\
 \cmidrule(lr){2-2}  \cmidrule(lr){3-3}  \cmidrule(lr){4-4} \cmidrule(lr){5-5} \cmidrule(lr){6-6} 
0 & 0.33 & 0.37 & 0.33 & 0.33 & 0.32\\
1 & 0.11 & 0.10 & 0.10 & 0.10 & 0.11\\
2 & 0.18 & 0.16 & 0.13 & 0.13 & 0.17\\
3 & 0.09 & 0.09 & 0.09 & 0.09 & 0.09\\
4 & 0.30 & 0.29 & 0.35 & 0.35 & 0.31\\
N & 2,507 & 827 & 804 & 905 & 11,534
\end{tabular}

\vspace{1ex}
}
\end{center}
  \caption{{\textbf{Shares of pre-test stimuli across timeline and Messenger by batch.} The sample is all users in the learning ($4,761$) and evaluation stage ($11,534$); this sample includes users who may have dropped out of the survey after prettest response was collected. Rows represent number of times respondents shared  true or false pre-test stimuli; columns represent batches; cells are the proportion of respondents within a respective batch that shared a type of stimuli a number of times. Batches 1-4 are in the learning stage; batch 5 is in the evaluation stage.}} \refstepcounter{SItable}\label{tab:shares}%
\end{table}

\clearpage

\subsection{Survey instrument}
The survey script is available at this link:\\
\if0\blind{%
\url{http://bit.ly/facebook_survey_public} }\fi
\if1\blind{%
[blinded] }\fi

All of the stimuli (posts) used in the experiment are available at this link:\\
\if0\blind{%
\url{http://bit.ly/facebook_stimuli_public} }\fi
\if1\blind{%
[blinded] }\fi

\subsection{Treatments}\label{appendix:treatments}

Treatments 1, 2, 3, 8, 9, and 10 are derived from interventions used by social media platforms including Facebook, Twitter, and WhatsApp. 
Treatment 11 (real information) is a similar headline-level treatment that \textit{could} be adopted by industry partners. Rather than flags or warnings about misinformation, we test whether providing a simple true statement reduces sharing of false information. Existing research suggests that providing true information can sometimes influence individuals' attitudes and behaviors \citep{gilens2001political}. Treatments 4, 6, and 7 are taken from previous academic studies. Emotions (4) have been suspected to influence susceptibility to misinformation \citep{martel2019reliance,rosenzweig2021happiness,bago2022emotion}; our test evaluates one canonical method of emotion suppression as a way to reduce the influence of misinformation. The accuracy nudge treatment (6) was specifically found to be effective at reducing the sharing of COVID-19 misinformation among participants in the US. Our deliberation nudge treatment (7) was adapted from ref. \citenum{bago2020fake} that found asking participants to deliberate was effective at improving discernment of online political information. The pledge treatment (5) was adapted from the types of treatments used by political campaigns to get subjects to pledge to vote or support a particular candidate \citep{costa2018walking}. We varied whether the pledge is made in private (within the chatbot conversation) or in public (posted on the respondent's Facebook timeline) to test whether public pledges are more effective at influencing behavior than private ones \citep{cotterill2013impact}. 

\begin{table}[H]
\centering
\resizebox{\textwidth}{!}{ 
\begin{tabular}{l|l|l}
\multicolumn{1}{l|}{\textbf{\begin{tabular}[c]{@{}c@{}}Shorthand\\ Name\end{tabular}}} & \multicolumn{1}{c|}{\textbf{\begin{tabular}[c]{@{}c@{}}Treatment\\ Level\end{tabular}}} & \textbf{Treatment}                                                                                                                                                                                                                                                                                                                                                                                              \\ \hline
1. Facebook tips                                                                                                           & Respondent                                                                                                   &  Facebook's ``Tips to Spot False News'' 
\\
2. AfricaCheck tips                                                                                                         & Respondent                                                                                                   &  \href{www.Africacheck.org}{Africacheck.org}'s guide: \\ & & ``How to vet information during a pandemic''                                                                                                                                                                                                                                                                                                                             \\
3. Video training
 & Respondent                                                                                                   &   \href{https://www.facebook.com/Vodcasts/videos/1322816708106278/}{BBC video} on spotting Coronavirus misinformation 
 \\
4. Emotion suppression                                                                                                       & Respondent                                                                                                   & \begin{tabular}[t]{@{}l@{}}Prompt: ``As you view and read the headlines, if you have any \\feelings, please try your best not to let those feelings show.  \\Read all of the headlines carefully, but try to behave so that \\someone watching you would not know that you are feeling\\ anything at all'' \citep{gross1998emerging}.\end{tabular}
\\
5. Pledge                                                                                 & Respondent                                                                                                   &  \begin{tabular}[t]{@{}l@{}} Prompt: Respondents will be asked if they want to keep their\\ family and friends safe from COVID-19, if they knew \\COVID-19 misinformation can be dangerous, and if they're\\ willing to take  a \textit{public} pledge to help identify\\and call out COVID-19 misinformation online. 
\end{tabular}
\\
6. Accuracy nudge                                                                                 & Respondent                                                                                                   & Placebo headline: ``To the best of your knowledge, is this\\& &headline accurate?'' \citep{pennycook2020fighting, pennycook_epstein_mosleh_arechar_eckles_rand_2019}.
\\
7. Deliberation nudge                                                                                 & Respondent                                                                                                   & Placebo headline: ``In a few words, please say \textit{why} you would\\ & & or would not like to share this story on Facebook.''\\ & & [open text response]
\\
8. Related articles                                                                                                       & Headline                                                                                                     & Facebook-style related stories: below story, show one other\\ & &  story that corrects a false news story                                                                                                                                                                                                                                                                                             \\
9. Fact check                                                                                                      & Headline                                                                                                     & Indicates story is ``Disputed by 3rd party fact-checkers''
 \\
10. More information                                                                                                      & Headline                                                                                                     & Provides a message and link to ``Get the facts about COVID-19''\\
11. Real information                                                                                                      & Headline                                                                                                     & Provides a \textit{true} statement: ``According to the WHO,\\ & & there is currently \textbf{no proven} cure for COVID-19.''
 \\
12. Control                                                                                                        & N/A                                                                                                          & Control condition                                                                                                                                                                                                                                                                                                                                                                                              
\end{tabular}
}
\caption{\textbf{Full list of treatments run during the learning phase.}}
\refstepcounter{SItable}\label{tab:treatments}
\end{table}

\paragraph{Facebook Tips}\label{sec:fbtips}
The script for the Facebook tips respondent-level treatment is as follows:

As we're learning more about the Coronavirus, new information can spread quickly, and it's hard to know what information and sources to trust. Facebook has some tips for how to be smart about what information to trust. 

1. Be skeptical of headlines. False news stories often have catchy headlines in all caps with exclamation points. If shocking claims in the headline sound unbelievable, they probably are.

2. Look closely at the link. A phony or look-alike link may be a warning sign of false news. Many false news sites mimic authentic news sources by making small changes to the link. You can go to the site to compare the link to established sources.

3. Investigate the source. Ensure that the story is written by a source that you trust with a reputation for accuracy. If the story comes from an unfamiliar organization, check their ``About'' section to learn more.

4. Watch for unusual formatting. Many false news sites have misspellings or awkward layouts. Read carefully if you see these signs.

5. Consider the photos. False news stories often contain manipulated images or videos. Sometimes the photo may be authentic, but taken out of context. You can search for the photo or image to verify where it came from.

6. Inspect the dates. False news stories may contain timelines that make no sense, or event dates that have been altered.

7. Check the evidence. Check the author's sources to confirm that they are accurate. Lack of evidence or reliance on unnamed experts may indicate a false news story.

8. Look at other reports. If no other news source is reporting the same story, it may indicate that the story is false. If the story is reported by multiple sources you trust, it's more likely to be true.

9. Is the story a joke? Sometimes false news stories can be hard to distinguish from humor or satire. Check whether the source is known for parody, and whether the story's details and tone suggest it may be just for fun.

10. Some stories are intentionally false. Think critically about the stories you read, and only share news that you know to be credible.

\subsection{Covariates}
\label{section:covariates}

In all analyses, we include the pretest response strata for true and false stimuli. For some continuous covariates that describe individual characteristics, such as education, we include an indicator flag if the respondent skipped the question; this is noted in the ``Coded as'' column. For others which require reflection or where there is a ``correct'' or ``best'' response, such as the Cognitive Reflection Test or the COVID-19 information measure, we code the index as 0 if the respondent chose not to answer any of the questions.

\begin{table}[H]
\begin{adjustbox}{totalheight=.9\textheight-2\baselineskip, max width = \textwidth}
\begin{tabular}{p{0.3\linewidth}p{0.7\linewidth}p{0.25\linewidth}}
\textbf{Covariate}                   & \textbf{Response options} & \textbf{Coded as}                                     \\
\hline
Gender                                      & Male,   Female, Nonbinary, Other                           & 1 if male, 0 otherwise  \\
Age                                         & Integers                                                   & Continuous, {flag if greater than 120}              \\
Education &
  No   formal schooling, Informal schooling only, Some primary school, Primary   school completed, Some secondary school, Secondary school completed,   Post-secondary qualifications, Some university, University completed,   Post-graduate &
  1:10, flag if missing \\
Geography                                   & Urban, Rural                                 & 1 if urban, 0 otherwise \\
Religion                                    & Christian, Muslim, Other/None                           & Indicators              \\
Denomination (Christian)  & Pentecostal, Other  & Indicator (coded 1 if Pentecostal, 0 otherwise)\\
Religiosity   (freq. of attendance) &
  Never,   Less than once a month, One to three times per month, Once a week, More than   once a week but less than daily, Daily &
  1:6, flag if missing \\
 Locus of control & 
[See survey instrument for full list] & 1:10, flag if missing\\
Index   of scientific views                 & [See   survey instrument for full questions and response options] & 0:2, flag if missing                     \\
Digital Literacy Index &  {[}Based on the first nine items of \cite{guessetal2020digital}'s  proposed measure, see  survey instrument for full questions and response options{]}& 0:24\\
Frequency of social media usage (x2)& {[}See   survey instrument for full questions and response options{]} & 0:3, flag if missing \\
Cognitive Reflection Test& {[}See   survey instrument for full questions and response options{]}& 0:3 (1 point for each correct response)\\
Index of household possessions
&
  I/my household owns, Do not own [See survey instrument for items] &
  Continuous, sum of owned items, flag if all missing \\
Job   with cash income                      & Yes,   No                                                  & 1 if yes                \\
Number   of people in household             & Integers                                                   & Continuous, flag if missing              \\
Political affiliation & Governing party v. opposition & Indicator (coded 1 if associate with or voted for candidate from governing party, 0 otherwise)\\
Concern regarding COVID-19                  & Not at all worried, Somewhat worried,  Very   worried      & 1:3, flag if missing                     \\
Perceived government efficacy   on COVID-19 & Very   poorly, Somewhat poorly, Somewhat well, Very well   & 1:4, flag if missing \\
{Strata of response to pre-test stimuli} & [Would share stimuli on timeline/via Messenger]& Indicators for strata (0:2) x (True + False = 2 types) $\times$ (timeline + Messenger = 2 channels) 
\end{tabular} 
\end{adjustbox}
\caption{\textbf{Covariates and response options.} Regarding missingness flags, respondents must respond to chatbot questions to advance in the survey, but for contexts they may enter ``skip'' if they do not wish to answer a given question, with the exception of age, which we check is greater than 18. }
\refstepcounter{SItable}\label{tab:cov_long}
\end{table}

\subsection{Response measurement}

We  are  primarily  interested  in  decreasing  sharing  of  harmful  false  information  about COVID-19 cures and treatments, but we simultaneously wish to limit any negative impact on sharing of useful information about transmission and best practices from verified sources.  In this case,  we care more about the spread of false COVID cures because in an environment of fear and uncertainty, belief that a cure will work may not play a large role in whether an individual tries a particular treatment when no proven alternative exists. We measure sharing intentions with two questions asked after each post the user saw: 1) would you like to share this post on your timeline? 2) would you like to send this post to a friend on Messenger?

\begin{figure}[H]
\centering
\begin{tikzpicture}[node distance=1.5cm,
    every node/.style={fill=white, font=\sffamily}, align=center]
 \node (pre)      [draw=none,rectangle, fill=none]
                                                      {\ \ $\cdots$ };    
 \node (pretreat)             [activityStarts, below of = pre]              {Pre-Treatment Stimuli \\ \small{[Random: 2 true/2 false]}};
  \node (distract)      [process, below of=pretreat, yshift=-.7in]
                                                      {Intermediate Modules};
  \node (treat)      [activityRuns, below of=distract]
                                                      {Treatment};
  \node (posttreat)      [activityStarts, below of=treat]
                                                      {Post-Treatment Stimuli \\ \small{[Random: 2 true/2 false]}};  
  \node (etc)      [below of=posttreat, yshift=-.7in, draw=none,rectangle, fill=none]
                                                      {\ \ $\cdots$ };    
    \draw[->]             (pre) -- (pretreat);    
     \draw[->]             (pretreat) -- node[text width=4.5in]
     				{
                                                $M_i^a, T_i^a
                                                \left\lbrace
                                                \begin{tabular}{ l} 
                                                1. Would you like to share this post on your timeline?\\
                                                2. Would you like to send this post to a friend?
                                                \end{tabular}\right.
                                                $
						 }(distract);
     \draw[->]             (distract) -- (treat);    
     \draw[->]             (treat) -- (posttreat);       
     \draw[->]             (posttreat) -- (etc);                      
     \draw[->]             (posttreat) -- node[text width=4.5in]
                                   {
                                                $M_i^b, T_i^b 
                                                \left\lbrace
                                                \begin{tabular}{ l} 
                                                1. Would you like to share this post on your timeline?\\
                                                2. Would you like to send this post to a friend?
                                                \end{tabular}\right.
                                                $
                                                }(etc);
\end{tikzpicture}
\caption{\textbf{Survey flow.}}
\refstepcounter{SIfig}\label{fig:survey_flow}
\end{figure}

We code responses to the self-reported questions as one if the respondent affirms they want to share the post and zero otherwise. Let $M_i^a$ be the sum of respondent $i$'s pretest responses to the \textit{misinformation} stimuli and let $T_i^a$ be the sum of respondent $i$'s pretest responses to the \textit{true} informational stimuli. We denote the respective sums of post-treatment responses by $M_i^b$ and $T_i^b$. By construction, $M_i^a, T_i^a, M_i^b, T_i^b \in \{0,1,2, 3, 4\}$. 

We formalize our response function in terms of posttest measures:
\[
Y_i = -M^b_i + 0.5 T^b_i.
\]
This response function is the metric for which we optimize in our adaptive algorithm. Table \ref{tab:response_fun} illustrates the values this discernment measure could take based on the number of intended true and false shares.

\begin{table}[!ht]
\centering
\begin{tabular}{lrrrrrr}
                                   & & \multicolumn{5}{c}{\textbf{True shares}}     \\ 
                                   & & 0                    & 1                    & 2                                 & 3                    & \multicolumn{1}{r}{4}    \\  \cmidrule(lr){3-7}
\multicolumn{1}{l}{}              & \multicolumn{1}{r}{0} & 0.0                  & 0.5                  & 1.0                               & 1.5                  & \multicolumn{1}{r}{2.0}  \\
\multicolumn{1}{l}{}              & \multicolumn{1}{r}{1} & $-1.0$                 & $-0.5$                 & 0.0                               & 0.5                  & \multicolumn{1}{r}{1.0}  \\
\multicolumn{1}{l}{\textbf{False shares}} & \multicolumn{1}{r}{2} & $-2.0$                 & $-1.5$                 & $-1.0$                              & $-0.5$                 & \multicolumn{1}{r}{0.0}  \\
\multicolumn{1}{l}{}              & \multicolumn{1}{r}{3} & $-3.0$                 & $-2.5$                 & $-2.0$                              & $-1.5$                 & \multicolumn{1}{r}{$-1.0$} \\
\multicolumn{1}{l}{}              & \multicolumn{1}{r}{4} & $-4.0$                 & $-3.5$                 & $-3.0$                              & $-2.5$                 & \multicolumn{1}{r}{$-2.0$} \\ 
\end{tabular}
\caption{\textbf{Discernment measure.}}
\refstepcounter{SItable}\label{tab:response_fun}
\end{table}

\section{Supplementary Results}
\subsection{Learning stage}\label{appendix:learning}%

\begin{figure}[H]
\centering
\begin{subfigure}[b]{0.5\textwidth}
        \centering
        \includegraphics[width = 0.95\textwidth]{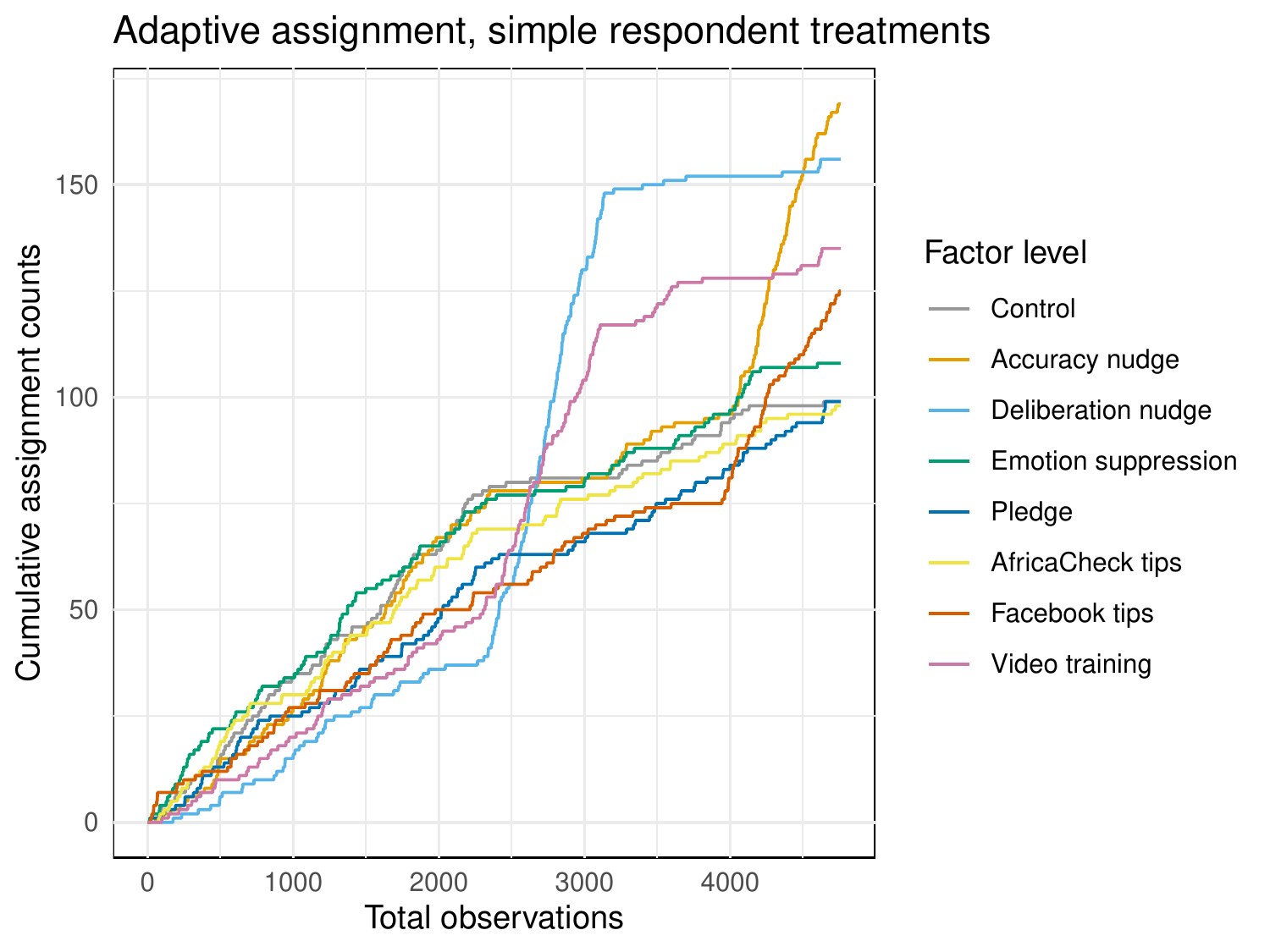}
    \end{subfigure}%
    \begin{subfigure}[b]{0.5\textwidth}
        \centering
        \includegraphics[width = 0.95\textwidth]{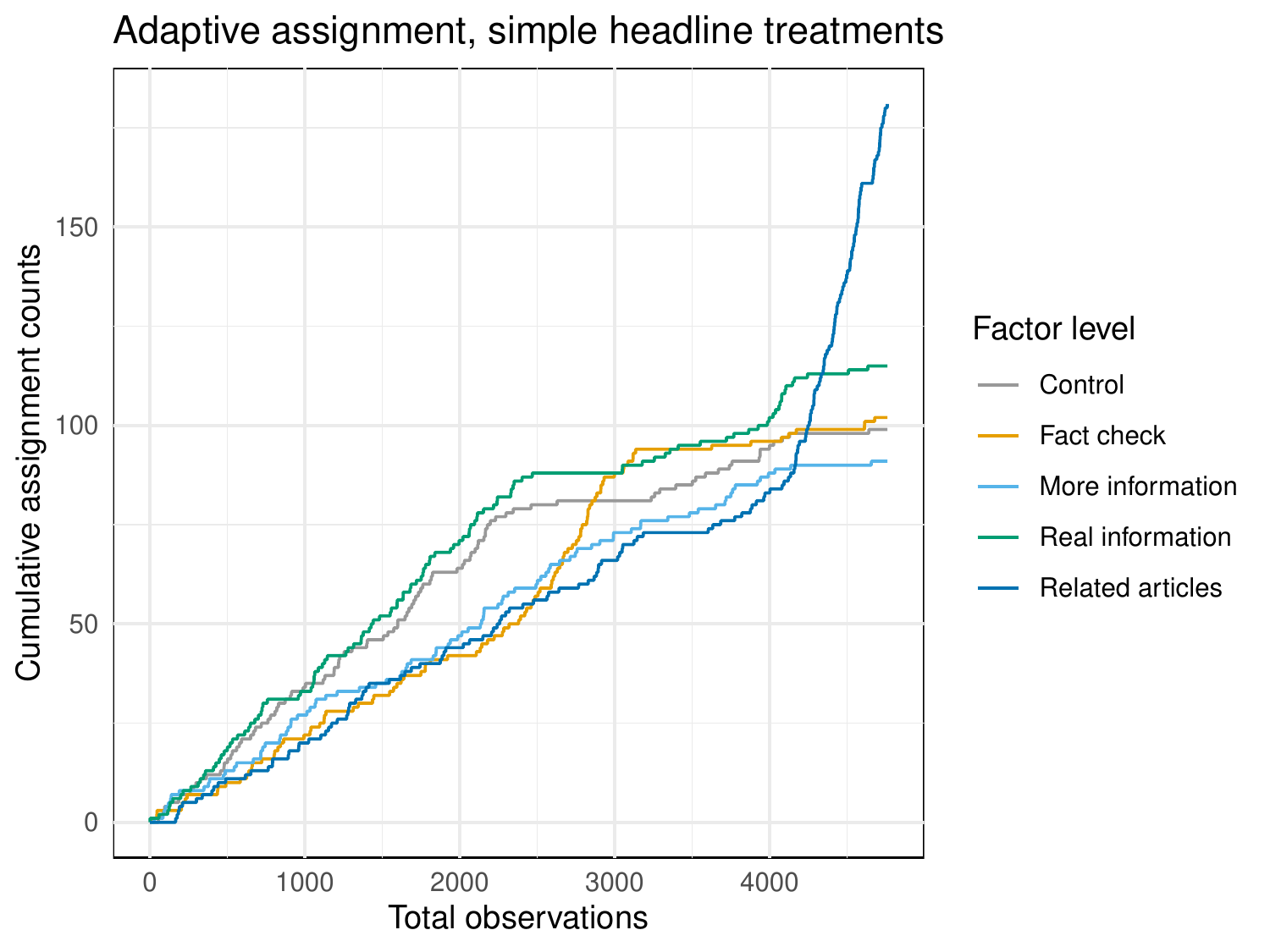}
    \end{subfigure}
   \caption{\textbf{Cumulative treatment assignment during the learning phase for respondent (left panel) and headline (right panel) interventions.} The sample is users in the learning stage, total $n = 4,761$. While the full design allows for all factor combinations, these plots illustrate cumulative assignment using only the ``simple'' version of each factor, i.e., when the other factor is at the baseline control condition.}
   \refstepcounter{SIfig}\label{fig:learning_cumulative_pure}
\end{figure}

{We illustrate learning stage assignment in \autoref{fig:learning_cumulative_pure}. Our algorithm is described in further detail in our pre-registration, along with justification for algorithm hyper-parameter selection. However, we note that the version of Balanced Linear Thompson Sampling that we use follows the algorithm described in ref. \citenum{dimakopoulou2017estimation} closely.}

{Our adaptive algorithm updates over four batches. In the first batch, treatment is assigned uniformly at random. The first batch is designed to be largest, with about 2,300 observations, so that the algorithm will have sufficient data to update, and so that assignment probabilities will be more stable. Following the first batch, the algorithm updates approximately every 800 observations. }

The adaptive assignment privileges assignment to those interventions that are predicted to be most effective, down-weighting assignment to interventions that are predicted to perform poorly. This means that we collect more data about the interventions that are the most likely to succeed. It is important to note that adaptively collected data introduces additional challenges for policy learning \citep{zhan2022policy}; the exploitation of the algorithm can eventually result in extreme probabilities of treatment assignment. However, this exploitation is an important ethical consideration in a setting where we are concerned about avoiding ``backfire'' from counter-productive interventions. The adaptive algorithm allows us to minimize these potentially harmful effects. {In our setting, we include probability floors of 1/400 to ensure that no treatment conditions are dropped entirely from the experiment.}

{\autoref{fig:learning_cumulative_pure} shows cumulative assignment to respondent treatments (left panel) when the headline factor level is fixed at control; and headline treatments (right panel) when the respondent factor level is fixed at control. We note that we use a ridge outcome model for adaptive assignment, which is different from the forest outcome model used for estimates presented in 
Figure 1, and so the policies selected as ``best'' by the adaptive algorithm need not be those with the highest estimates ex-post; with sufficient data, however, differences should be small. We can see in the left panel that the algorithm selected the deliberation nudge and the video training as promising candidates around 2,300 observations, after the first batch, as assignment to these two conditions is rapidly increasing; however assignment to these treatments flattens out shortly after 3,000 observations. By the last batch, after 4,000 observations, assignment to the accuracy nudge and Facebook tips is most rapidly increasing, indicating that the algorithm is assigning these treatments with the highest probabilities. In the right panel, it is not evident that any treatment has taken a clear lead until the last batch, when there is an uptick in assignment to the related articles treatment.}

{In general it appears that the ``best'' arms learned by the algorithm by the end of the adaptive experiment are consistent with those selected for the evaluation stage from our estimates presented in 
Figure 1. However, for illustration purposes we have only included in \autoref{fig:learning_cumulative_pure} the ``simple'' version of each factor, i.e., when the other factor is at the baseline control condition. To account for the contextual nature of the algorithm, we consider in  \autoref{tab:learning_cumulative} how assignment under the algorithm matches our Restricted Targeted Policy, estimated on all of the learning stage data. In calculating the probability of assignment to the this policy, we ignore headline-level assignment (which was also randomized in the factorial design), since the Learned and Restricted Targeted Policies vary only the respondent level treatment. We see that over time, the share of units whose learning-stage assignment matches the Restricted Targeted Policy increases.}

\begin{table}[H]
\centering

\begin{tabular}[t]{lcc}
 & \textbf{Learned}& \textbf{Restricted} \\ 
 \cmidrule(lr){2-2} \cmidrule(lr){3-3}
Batch 1 & \num{0.125} & \num{0.125}\\
 & (\num{0.000}) & (\num{0.000})\\
Batch 2 & \num{0.153} & \num{0.154}\\
 & (\num{0.002}) & (\num{0.002})\\
Batch 3 & \num{0.147} & \num{0.153}\\
 & (\num{0.002}) & (\num{0.002})\\
Batch 4 & \num{0.257} & \num{0.255}\\
 & (\num{0.004}) & (\num{0.004})\\
\end{tabular}

\caption{{\textbf{Evolution of on-policy probabilities during the learning stage.} The sample is users in the learning stage, total $n = 4,761$. Rows are share of participants in each batch assigned to the respondent-level treatment they would be assigned under the Learned Targeted Policy or the Restricted Targeted Policy. }}
\refstepcounter{SItable}\label{tab:learning_cumulative}
\end{table}

\begin{table}[H]
   \centering
\resizebox{\textwidth}{!}{ 

\begin{tabular}[t]{lccccc}
 & \textbf{Control} & \textbf{Fact check} & \textbf{More information} & \textbf{Real information} & \textbf{Related articles}\\ 
 \cmidrule(lr){2-2} \cmidrule(lr){3-3}\cmidrule(lr){4-4}\cmidrule(lr){5-5}\cmidrule(lr){6-6}
Control & \num{-0.334} & \num{-0.082} & \num{0.077} & \num{0.127} & \num{-0.119}\\
 & (\num{0.236}) & (\num{0.184}) & (\num{0.195}) & (\num{0.165}) & (\num{0.125})\\
Accuracy nudge & \num{0.189} & \num{0.044} & \num{0.103} & \num{0.022} & \num{-0.032}\\
 & (\num{0.204}) & (\num{0.097}) & (\num{0.162}) & (\num{0.156}) & (\num{0.119})\\
Deliberation nudge & \num{-0.155} & \num{-0.294} & \num{-0.431} & \num{-0.050} & \num{0.335}\\
 & (\num{0.192}) & (\num{0.144}) & (\num{0.256}) & (\num{0.153}) & (\num{0.168})\\
Emotion suppression & \num{0.063} & \num{-0.110} & \num{0.228} & \num{-0.139} & \num{0.013}\\
 & (\num{0.144}) & (\num{0.137}) & (\num{0.120}) & (\num{0.143}) & (\num{0.171})\\
Pledge & \num{0.070} & \num{0.214} & \num{-0.101} & \num{-0.225} & \num{-0.001}\\
 & (\num{0.189}) & (\num{0.333}) & (\num{0.142}) & (\num{0.139}) & (\num{0.120})\\
AfricaCheck tips & \num{0.376} & \num{-0.005} & \num{0.024} & \num{-0.197} & \num{-0.312}\\
 & (\num{0.390}) & (\num{0.131}) & (\num{0.130}) & (\num{0.133}) & (\num{0.208})\\
Facebook tips & \num{-0.074} & \num{0.096} & \num{-0.095} & \num{0.123} & \num{0.107}\\
 & (\num{0.132}) & (\num{0.164}) & (\num{0.221}) & (\num{0.235}) & (\num{0.223})\\
Video training & \num{-0.254} & \num{0.082} & \num{0.121} & \num{0.075} & \num{-0.099}\\
 & (\num{0.145}) & (\num{0.141}) & (\num{0.168}) & (\num{0.143}) & (\num{0.142})\\
\end{tabular}

}
\caption{{\textbf{Estimation of interaction effects over group means in the learning stage; varying respondent, fixing headline.} The sample is users in the learning stage, total $n = 4,761$. 
Rows are respondent-level treatments, columns are headline-level treatments. Estimates are in terms of discernment, and represent differences in effects between the row $\times$ column interaction effect and the column effect, averaging over rows. Estimates are produced from differences in adaptively weighted augmented inverse probability weighted estimators, as described in \autoref*{section:estimation}. 
}}
\refstepcounter{SItable}\label{tab:learning_interaction_headline}
\end{table}

\begin{table}[H]
   \centering
\resizebox{\textwidth}{!}{ 

\begin{tabular}[t]{lcccccccc}
 & \textbf{Control} & \textbf{Accuracy nudge} & \textbf{Deliberation nudge} & \textbf{Emotion suppression} & \textbf{Pledge} & \textbf{AfricaCheck tips} & \textbf{Facebook tips} & \textbf{Video training}\\ 
 \cmidrule(lr){2-2} \cmidrule(lr){3-3}\cmidrule(lr){4-4}\cmidrule(lr){5-5}\cmidrule(lr){6-6}\cmidrule(lr){7-7}\cmidrule(lr){8-8}\cmidrule(lr){9-9}
Control & \num{-0.294} & \num{0.065} & \num{-0.055} & \num{0.038} & \num{0.083} & \num{0.392} & \num{-0.158} & \num{-0.240}\\
 & (\num{0.237}) & (\num{0.202}) & (\num{0.195}) & (\num{0.140}) & (\num{0.190}) & (\num{0.395}) & (\num{0.136}) & (\num{0.142})\\
Fact check & \num{-0.001} & \num{-0.039} & \num{-0.152} & \num{-0.093} & \num{0.268} & \num{0.052} & \num{0.053} & \num{0.137}\\
 & (\num{0.189}) & (\num{0.099}) & (\num{0.152}) & (\num{0.136}) & (\num{0.335}) & (\num{0.149}) & (\num{0.170}) & (\num{0.142})\\
More information & \num{0.036} & \num{-0.102} & \num{-0.411} & \num{0.123} & \num{-0.169} & \num{-0.040} & \num{-0.260} & \num{0.055}\\
 & (\num{0.200}) & (\num{0.164}) & (\num{0.261}) & (\num{0.121}) & (\num{0.149}) & (\num{0.149}) & (\num{0.226}) & (\num{0.170})\\
Real information & \num{0.187} & \num{-0.082} & \num{0.070} & \num{-0.144} & \num{-0.193} & \num{-0.162} & \num{0.058} & \num{0.109}\\
 & (\num{0.173}) & (\num{0.160}) & (\num{0.162}) & (\num{0.145}) & (\num{0.148}) & (\num{0.153}) & (\num{0.241}) & (\num{0.146})\\
Related articles & \num{-0.031} & \num{-0.108} & \num{0.484} & \num{0.036} & \num{0.060} & \num{-0.248} & \num{0.071} & \num{-0.037}\\
 & (\num{0.134}) & (\num{0.122}) & (\num{0.176}) & (\num{0.172}) & (\num{0.129}) & (\num{0.220}) & (\num{0.229}) & (\num{0.144})\\
\end{tabular}

}
\caption{{\textbf{Estimation of interaction effects over group means in the learning stage; varying headline, fixing respondent.} The sample is users in the learning stage, total $n = 4,761$. 
Rows are headline-level treatments, columns are respondent-level treatments. Estimates are in terms of discernment, and represent differences in effects between the row $\times$ column interaction effect and the column effect, averaging over rows. Estimates are produced from differences in adaptively weighted augmented inverse probability weighted estimators, as described in \autoref*{section:estimation}. 
}}
\refstepcounter{SItable}\label{tab:learning_interaction_respondent}
\end{table}


\clearpage

\subsection{Evaluation stage}\label{appendix:evaluation}




\subsubsection{{Covariate variation in original policy}}

\begin{figure}[H] 
   \centering
   \includegraphics[width=\textwidth]{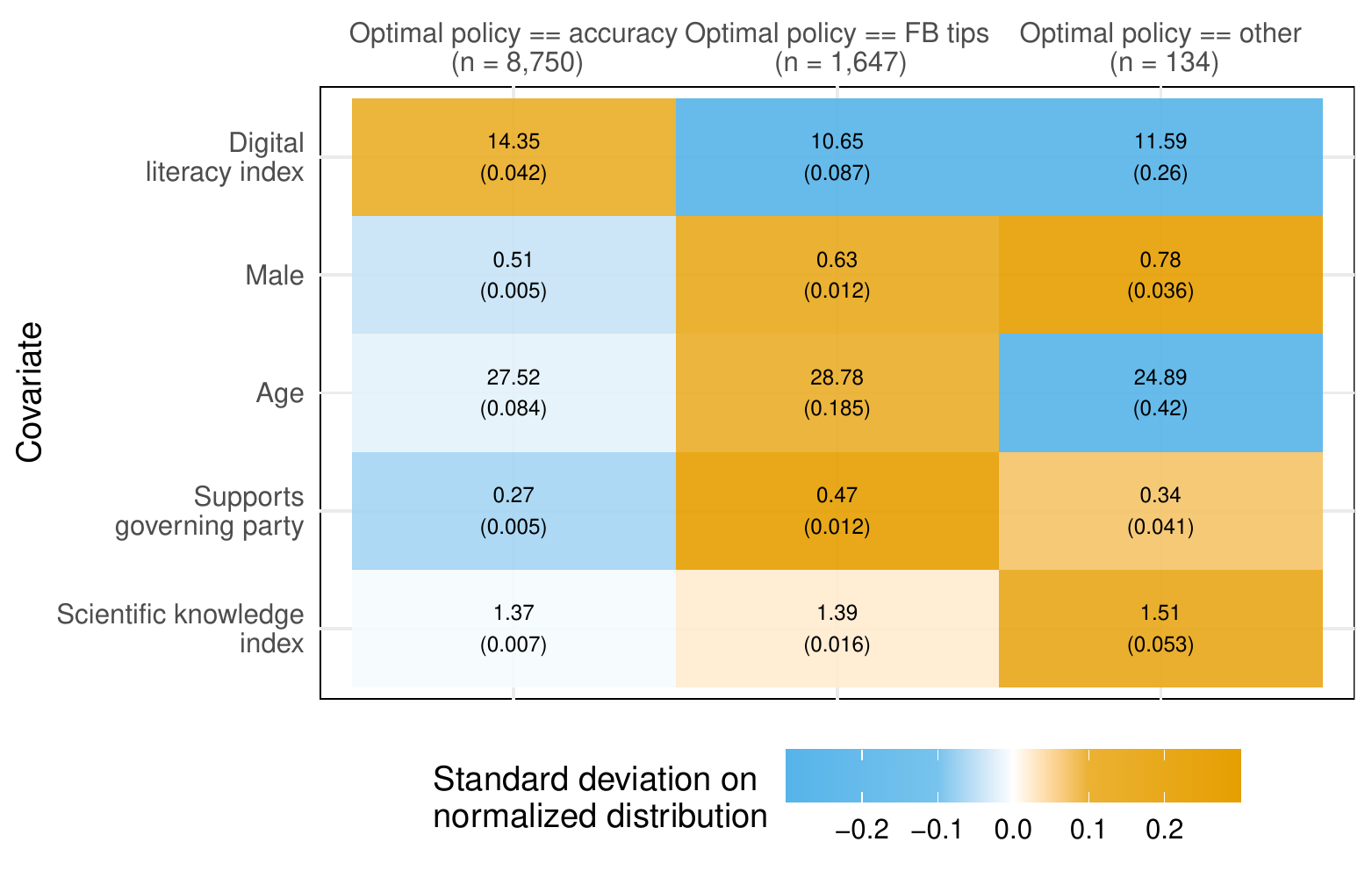} 
   \caption{\textbf{Selected covariate means by Learned Targeted Policy assignment.} The sample is users in the evaluation stage, $n = 10,531$. Covariates are ordered by size of standardized deviation between the first two groups. }
   \refstepcounter{SIfig}\label{fig:heterogeneity_covariates_original}
\end{figure}

In \autoref{fig:heterogeneity_covariates_original}, we report differences in selected covariates across the groups assigned to each treatment in the Learned Targeted Policy. 
The participants assigned to the \textit{accuracy nudge} are, on average, more digitally literate and less likely to be male than participants assigned to \textit{Facebook tips} or other respondent-level conditions.

\subsubsection{Benefits to personaliation: Rank-weighted average treatment effects}
\begin{figure}[H] 
   \centering
   \includegraphics[width=\textwidth]{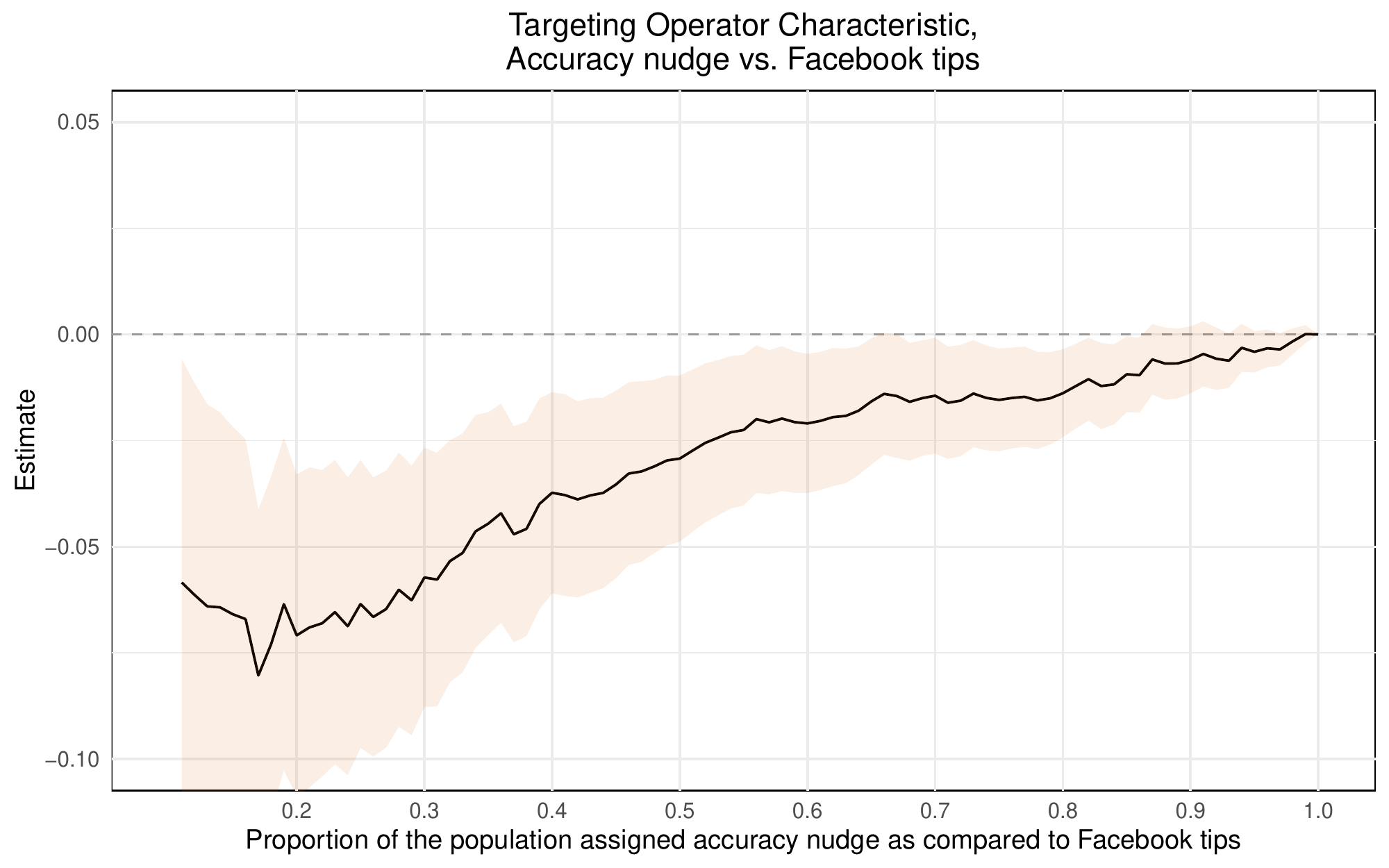} 
   \caption{\textbf{Targeting operator characteristic curve, comparing the accuracy nudge and Facebook tips.} The policy is estimated using the learning stage data. The sample for evaluation here is users in the evaluation stage, $n = 10,531$. The outcome measure is the difference in proportion of false stimuli participants reported wanting to share, either as a Facebook post or privately in Facebook Messenger, between the accuracy nudge and Facebook tips. The $y$-axis represents average differences in this measure if the users receiving the accuracy nudge were assigned according to a prioritization rule, as compared to at random. The shaded region shows the 95\% confidence interval.}
   \refstepcounter{SIfig}\label{fig:RATE}
\end{figure}

Ref. \citenum{yadlowsky2021evaluating} provide another approach to evaluate the benefits of personalization. Supposing hypothetically that a prespecified fraction of participants are to be allocated to the \textit{accuracy nudge} rather than \textit{Facebook tips}, we develop a targeted prioritization rule (following the same method for estimating counterfactual outcomes used to estimate the Restricted Targeted Policy) for allocating participants to the \textit{accuracy nudge}. 
We compare expected outcomes under this prioritization rule to the case where the same fraction of participants are allocated to \textit{accuracy nudge}, but participants are selected randomly. 
\autoref{fig:RATE} illustrates this benefit to targeting, as we vary the percentage of participants allocated to \textit{accuracy nudge}. 
If we were limited to assigning the \textit{accuracy nudge} to only 30 percent of the population and assigned \textit{Facebook tips} to the remainder, false sharing intentions would be 5.7~pp lower (s.e. = $1.6$, $Z = -3.67$, $p < 0.001$, 95\% CI = [$-8.77$, $-2.67$]) than if we had used the prioritization rule instead of random assignment. The overall rank-weighted average treatment effect, a weighted sum of the area under the curve in \autoref{fig:RATE}, is $-3.7$~pp (s.e. = $1.1$, $Z = -3.47$, $p < 0.001$, 95\% CI = [$-5.73$, $-1.59$]), using the targeting operator characteristic curve. 

\subsubsection{{Alternative specifications}}

{We report alternative specifications, first, of the estimation strategy for treatment effect estimates on sharing discernment.}
{
In \autoref{tab:alternative_results}, 
\begin{itemize}
\item The first five columns treat alternative covariate adjustment specifications. 
\begin{itemize}
\item Column (1) reports estimates produced from scores from an augmented inverse probability weighted estimator, using all covariates except pre-test response strata. This is the same estimating strategy used in our main result in \autoref*{tab:main_results}, but excluding pre-test response strata in the controls. We weight for treatment assignment probability and to account for attrition. 
\item Column (2) reports estimates produced from scores from an inverse probability weighted estimator, with no covariates. This is similar to the strategy used in our main result in \autoref*{tab:main_results}, but because we include no covariates, we do not have a conditional means model, and only weight for treatment assignment probability and to account for attrition. 
\item Column (3) reports estimates from a linear model with robust standard errors, with inverse probability treatment of treatment weights and no covariate adjustment. 
\item Column (4) reports estimates from a Lin estimator, with inverse probability of treatment weights, adjusting for continuous pre-test response. 
\item Column (5) reports estimates from a Lin estimator, with inverse probability of treatment weights, adjusting for strata of assignment based on the original Learned Targeted Policy.
\end{itemize}
\item The last three columns treat alternative approaches to accounting for missing data or covariate drift. 
\begin{itemize}
\item Column (6) reports estimates produced from scores from an augmented inverse probability weighted estimator, and all covariates. 
\item Column (7) reports estimates produced from scores from an augmented inverse probability weighted estimator, using all respondents assigned treatment, with missing post-test responses imputed from pre-test responses, and all covariates. 
\item Column (8) reports estimates produced from scores from an augmented inverse probability weighted estimator, with sample weights re-weighting to the distribution of covariates in the learning stage, and all covariates. 
\end{itemize}
\end{itemize}
}

{Augmented inverse probability weighted estimators are estimated using the form in \autoref*{section:estimation}, with subsets of covariates taking the form described in \autoref{tab:cov_long}, with the conditional means model estimated using a random forest. All specifications account for different treatment assigment probabilities. Specifications (1) and (2) also use estimated weights to account for censoring, for comparability with the main results in \autoref*{tab:main_results}. }

\begin{table}[H]
\centering
\resizebox{\textwidth}{!}{ 

\begin{tabular}[t]{lcccccccc}
 & (1) & (2) & (3) & (4) & (5) & (6) & (7) & (8)\\& \multicolumn{5}{c}{Alternative covariate adjustment} & \multicolumn{3}{c}{Missing data/ reweighting}\\ 
\cmidrule(lr){2-6} \cmidrule(lr){7-9} \multicolumn{4}{l}{\textbf{Headline treatment effects}} \rule{0pt}{1.2\normalbaselineskip}\\
\hspace{1em}Fact check & \num{-0.010} & \num{-0.027} & \num{-0.019} & \num{-0.051} & \num{-0.019} & \num{-0.030} & \num{-0.025} & \num{-0.033}\\
 & (\num{0.044}) & (\num{0.053}) & (\num{0.046}) & (\num{0.036}) & (\num{0.045}) & (\num{0.036}) & (\num{0.034}) & (\num{0.036})\\
\hspace{1em}Related articles & \num{-0.029} & \num{-0.045} & \num{-0.034} & \num{-0.059} & \num{-0.033} & \num{-0.047} & \num{-0.045} & \num{-0.046}\\
 & (\num{0.044}) & (\num{0.053}) & (\num{0.045}) & (\num{0.036}) & (\num{0.045}) & (\num{0.035}) & (\num{0.033}) & (\num{0.035})\\\multicolumn{4}{l}{\textbf{Respondent treatment effects}} \rule{0pt}{1.2\normalbaselineskip}\\
\hspace{1em}Accuracy & \num{0.091} & \num{0.102} & \num{0.080} & \num{0.060} & \num{0.081} & \num{0.068} & \num{0.062} & \num{0.068}\\
 & (\num{0.039}) & (\num{0.046}) & (\num{0.041}) & (\num{0.033}) & (\num{0.041}) & (\num{0.032}) & (\num{0.030}) & (\num{0.032})\\
\hspace{1em}Facebook tips & \num{0.075} & \num{0.081} & \num{0.097} & \num{0.057} & \num{0.093} & \num{0.057} & \num{0.039} & \num{0.061}\\
 & (\num{0.043}) & (\num{0.051}) & (\num{0.046}) & (\num{0.037}) & (\num{0.045}) & (\num{0.036}) & (\num{0.034}) & (\num{0.036})\\
\hspace{1em}Learned Targeted Policy & \num{0.075} & \num{0.076} & \num{0.065} & \num{0.049} & \num{0.060} & \num{0.059} & \num{0.054} & \num{0.071}\\
 \hspace{1.5em}(maximizing sharing discernment) & (\num{0.038}) & (\num{0.046}) & (\num{0.040}) & (\num{0.032}) & (\num{0.040}) & (\num{0.032}) & (\num{0.030}) & (\num{0.032})\\
\hspace{1em}Restricted Targeted Policy & \num{0.132} & \num{0.138} & \num{0.117} & \num{0.110} & \num{0.117} & \num{0.101} & \num{0.094} & \num{0.101}\\
 \hspace{1.5em}(minimizing any false sharing) & (\num{0.040}) & (\num{0.047}) & (\num{0.042}) & (\num{0.034}) & (\num{0.042}) & (\num{0.033}) & (\num{0.031}) & (\num{0.033})\\\hline
\hspace{1em}Control mean & \num{-0.423} & \num{-0.455} & \num{-0.419} & \num{-0.395} & \num{-0.419} & \num{-0.406} & \num{-0.419} & \num{-0.404}\\
 & (\num{0.031}) & (\num{0.037}) & (\num{0.032}) & (\num{0.026}) & (\num{0.032}) & (\num{0.027}) & (\num{0.025}) & (\num{0.027})\\\hline
\hline

{\textbf{Covariates}} \rule{0pt}{1.2\normalbaselineskip}\\

\hspace{1em} All & $-$ & $-$ & $-$ &  $-$ &  $-$ & Yes & Yes & Yes \\

\hspace{1em} Pre-test response only & $-$  & $-$& $-$ & Yes &  $-$ &  $-$ &  $-$ &  $-$ \\

\hspace{1em} Learned policy strata only & $-$ & $-$ &  $-$ &  $-$ &  Yes &  $-$ &  $-$ & $-$  \\

\hspace{1em} None & $-$ &Yes &  Yes &$-$ &   $-$ &  $-$ &  $-$ &  $-$ \\

{\textbf{Missing data}} \rule{0pt}{1.2\normalbaselineskip}\\

\hspace{1em} Imputing pre-test response & $-$ & $-$ &  $-$ &  $-$ &  $-$ &  $-$ & Yes &  $-$  \\

{\textbf{Weights}} \rule{0pt}{1.2\normalbaselineskip}\\

\hspace{1em} Treatment assignment probability &Yes &Yes  & Yes & Yes & Yes & Yes & Yes & Yes \\

\hspace{1em} Learning stage covariate distribution & $-$ & $-$ &  $-$ &  $-$ & $-$ &  $-$ &  $-$ &  Yes \\

\hspace{1em} Censoring &Yes &Yes &  $-$ &  $-$ &  $-$ &  $-$ &  $-$ &  $-$ 
\\

{\textbf{Estimator}} \rule{0pt}{1.2\normalbaselineskip}\\

\hspace{1em} (A)IPW Scores & Yes & Yes &  $-$ &  $-$ &  $-$ & Yes & Yes & Yes\\

\hspace{1em} Robust OLS &  $-$ &  $-$ & Yes &  $-$ &  $-$ &  $-$ & $-$ &  $-$ \\

\hspace{1em} Lin Estimator &  $-$ &  $-$ &  $-$ & Yes & Yes &  $-$ & $-$ &  $-$ \\

{\textbf{n}} & 10,531 & 10,531 & 10,531 & 10,531 & 10,531 & 10,531 &11,534  & 10,531 
\rule{0pt}{1.2\normalbaselineskip}\\

\end{tabular}

\vspace{1ex}
}
\caption{\textbf{Combined response function; control response and treatment effect estimates under alternative specifications.} The sample is users in the evaluation stage, $n = 10,531$, or $n = 11,534$ if including respondents with missing post-test response. Estimates are in terms of sharing discernment, a weighted sum of number of false sharing intentions (negatively
weighted) and true sharing intentions (positively weighted). 
The last row represents estimated mean response under the control condition; all other rows are estimated treatment effects in contrast with the control condition. 
Columns denote alternative specifications to those presented in \autoref*{tab:main_results}.
} 
\refstepcounter{SItable}\label{tab:alternative_results}
\end{table}

{We note here the benefits to controlling for pre-rest response; comparing our main, fully adjusted results in \autoref*{tab:main_results} to those in column (1) where we drop adjustment for pre-test response strata, the standard error on the control mean under sharing discernment decreases by 15.2\%. 
Similarly, under columns 3 and 4, when we move from our most simple linear model with no covariate adjustment, to one with controls for continuous false and true pre-test response, the standard error on the control mean is reduced by 20.7\%. While further covariate adjustment results in some changes to estimates, a great deal of the precision in adjusting for covariates is achieved by adjusting only for pre-test response.}

\subsubsection{{Alternative targeted policy: restricted treatments, targeted to combined response}}
{Below, we consider learning another contextual policy on the learning data, optimizing for the \textit{discernment} measure. This is similar to the approach reported in \autoref*{tab:main_results} for the Learned Targeted Policy, but here we constrain policy assignment to only the accuracy nudge or Facebook tips, as in the Restricted Targeted Policy, reported in that table. This policy's efficacy is similar to that under the original Learned Targeted Policy; we do not see great additional benefits to contextual policy learning when optimizing for the discernment measure. In this targeted policy, 66.3\% of respondents are assigned to the accuracy nudge, and 33.7\% to Facebook tips. There is 63.7\% overlap with the original Learned Targeted Policy, and 70.0\% overlap with the Restricted Targeted Policy.}

\begin{table}[H]
\centering
\resizebox{\textwidth}{!}{ 

\begin{tabular}[t]{lccccccc}
 & \textbf{Sharing} &  & \textbf{False} &  &  & \textbf{True} & \\
 & \textbf{Discernment} & Any sharing & Messenger & Timeline & Any sharing & Messenger & Timeline\\\cmidrule(lr){2-2} \cmidrule(lr){3-5} \cmidrule(lr){6-8}
\hspace{1em}Secondary targeted policy & \num{0.066} & \num{-0.028} & \num{-0.022} & \num{-0.026} & \num{-0.001} & \num{0.007} & \num{-0.003}\\
 \hspace{1.5em}(maximizing sharing discernment) & (\num{0.033}) & (\num{0.010}) & (\num{0.010}) & (\num{0.009}) & (\num{0.010}) & (\num{0.010}) & (\num{0.010})\\
\end{tabular}

\vspace{1ex}
}
\caption{\textbf{Control response and treatment effect estimates, policy learning with the combined response function.} The sample is users in the evaluation stage, $n = 10,531$. Columns denote response measures, described in the note to \autoref*{tab:heterogeneity_control}. 
Rows are estimated treatment effects in contrast with the control condition. Estimates are produced from an augmented inverse probability weighted estimator, as described in \autoref*{section:estimation}. 
} 
\refstepcounter{SItable}\label{tab:alternative_policy_results}
\end{table}

\subsubsection{{Alternative response measure: re-weighting}}
{Our discernment measure is a weighted sum of number of false sharing intentions (negatively
weighted) and true sharing intentions (positively weighted). For the main results reported in this paper, we use weights of -1 and 0.5 respectively. We consider here how significance of our estimates would have varied had we used an alternative weighting scheme, using response measures that are a linear combinations of weights. There is not a unique weighting scheme that provides the strongest tests across treatment conditions. In general, for the headline level treatments, allocating larger weights to false sharing results in larger negative test statistics. For the respondent-level treatments, there appears to be statistical benefit found in a more balanced combination.
The largest statistics are found between 0 and 0.5 on the logged negative ratio of false to true weights; 0 indicates parity in weights, e.g., -1 and 1 for false and true sharing respectively; 0.5 indicates weights of approximately -1 and 0.6. }

\begin{figure}[H] 
   \centering
   \includegraphics[width=\textwidth]{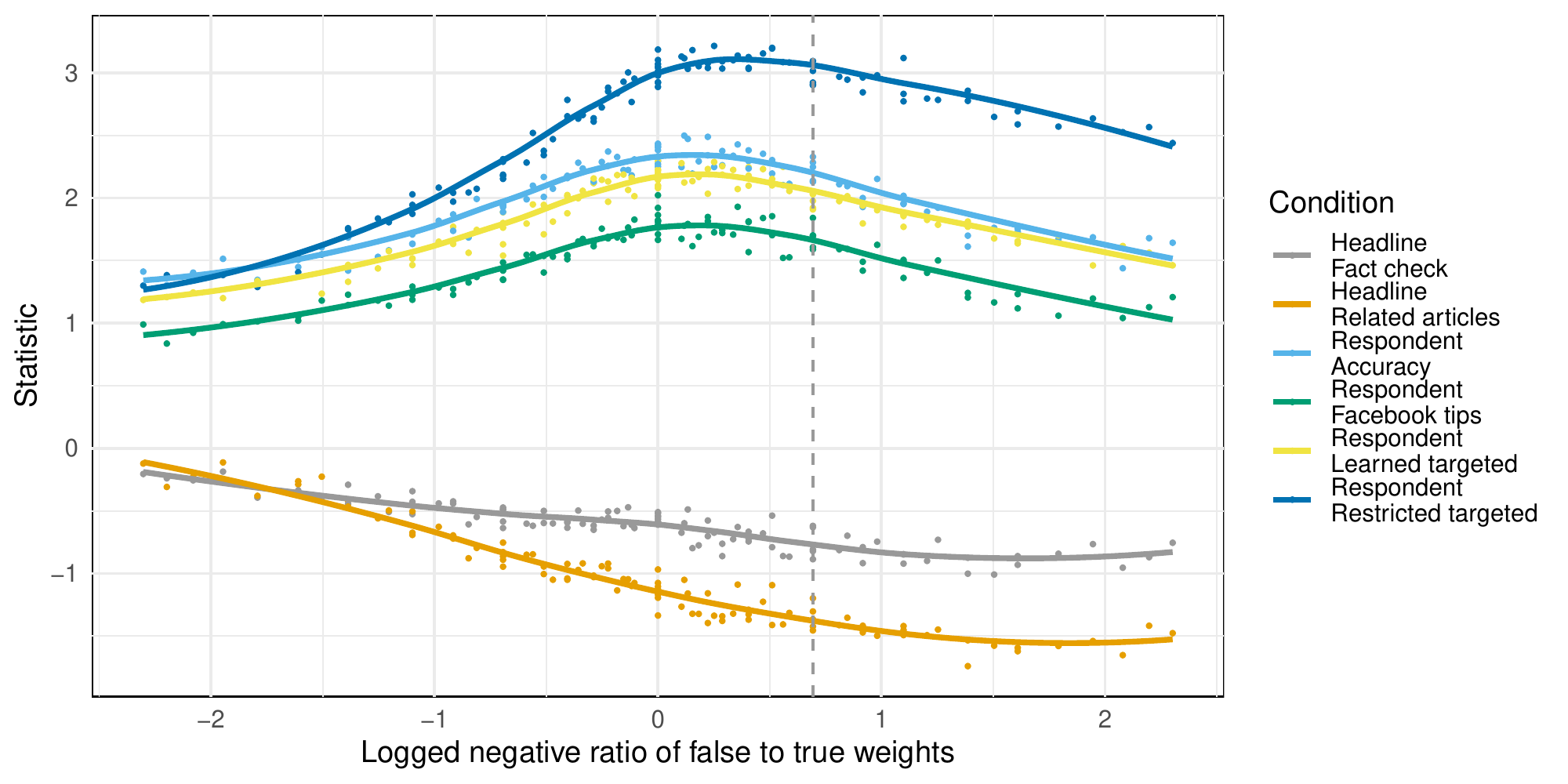} 
   \caption{\textbf{Significance of treatment effects on discernment under alternative weighting schemes.} The sample for evaluation here is users in the evaluation stage, $n = 10,531$. 
   The $y$-axis represents the t-statistic of the treatment effect estimate, the x-axis is the logged negative ratio of false to true weights. The dotted vertical line represents the transformed value for the weighting scheme used for this study, with weights of \num{-1} for false and \num{0.5} for true:  $\textrm{log}(2)$.
   We allow weights for false sharing to range from \num{-1} to \num{-0.1}; for true sharing from \num{0.1} to 1. Estimates are produced from re-calculating an augmented inverse probability weighted estimator, as described in \autoref*{section:estimation}, with the outcome weighted by different combinations of weights. }
   \refstepcounter{SIfig}\label{fig:alternative_weighting}
\end{figure}

\subsubsection{Heterogeneous response and treatment effects}

\label{appendix:heterogeneity}

Table \ref{tab:heterogeneity_control} illustrates that younger users, men, those aligned with the ruling party, participants with low digital literacy, and those with low scientific knowledge intend to share relatively more false stimuli under the control condition. For these ``worst offenders,'' we find that assigning the respondent-level treatments on average decreases false sharing as compared to control among men (estimate = $-3.3$~pp, s.e. = $1.3$, $Z = -2.57$, $p = 0.01$, 95\% CI = [$-5.76$, $-0.77$]), participants with low digital literacy (estimate = $-2.9$~pp, s.e. = $1.3$, $Z = -2.3$, $p = 0.022$, 95\% CI = [$-5.41$, $-0.43$]), and participants with low scientific knowledge (estimate = $-3.7$~pp, s.e. = $1.3$, $Z = -2.86$, $p = 0.004$, 95\% CI = [$-6.15$, $-1.15$]). (See \autoref*{tab:heterogeneity_treatment}.) The pooled respondent-level interventions do not reduce sharing of false posts among younger participants but do among older ones. Similarly, there is no effect of the pooled respondent treatments on false sharing among those aligned with the political party in power, but we do see a significant effect among those not aligned. However, \textit{differences} in treatment effects across groups are for the most part only statistically significant when comparing users with low to those with high levels of scientific knowledge.

\begin{table}[H]
\small
   \centering
\resizebox{0.95\textwidth}{!}{ 

\begin{tabular}[t]{lccccccc}
 & \textbf{Sharing} &  & \textbf{False} &  &  & \textbf{True} & \\
 & \textbf{Discernment} & Any sharing & Messenger & Timeline & Any sharing & Messenger & Timeline\\\cmidrule(lr){2-2} \cmidrule(lr){3-5} \cmidrule(lr){6-8} \multicolumn{4}{l}{\textbf{Age}} \rule{0pt}{1.2\normalbaselineskip}\\
\hspace{1em} Below median & \num{-0.004} & \num{-0.011} & \num{-0.012} & \num{-0.008} & \num{-0.018} & \num{-0.010} & \num{-0.020}\\
\hspace{2em}(n = 5,300) & (\num{0.040}) & (\num{0.013}) & (\num{0.013}) & (\num{0.012}) & (\num{0.013}) & (\num{0.013}) & (\num{0.013})\\
\hspace{1em} Above Median & \num{0.125} & \num{-0.031} & \num{-0.026} & \num{-0.033} & \num{0.031} & \num{0.031} & \num{0.024}\\
\hspace{2em}(n = 5,231) & (\num{0.045}) & (\num{0.013}) & (\num{0.013}) & (\num{0.013}) & (\num{0.013}) & (\num{0.013}) & (\num{0.013})\\\cmidrule(lr){2-8}
\hspace{1em} Difference & \num{0.128} & \num{-0.020} & \num{-0.014} & \num{-0.025} & \num{0.049} & \num{0.041} & \num{0.045}\\
\hspace{2em} & (\num{0.060}) & (\num{0.018}) & (\num{0.018}) & (\num{0.017}) & (\num{0.018}) & (\num{0.018}) & (\num{0.018})\\\multicolumn{4}{l}{\textbf{Gender}} \rule{0pt}{1.2\normalbaselineskip}\\
\hspace{1em} Not male & \num{0.037} & \num{-0.008} & \num{-0.010} & \num{-0.010} & \num{0.008} & \num{0.026} & \num{0.000}\\
\hspace{2em}(n = 4,915) & (\num{0.042}) & (\num{0.013}) & (\num{0.013}) & (\num{0.012}) & (\num{0.013}) & (\num{0.014}) & (\num{0.013})\\
\hspace{1em} Male & \num{0.080} & \num{-0.033} & \num{-0.027} & \num{-0.030} & \num{0.005} & \num{-0.003} & \num{0.003}\\
\hspace{2em}(n = 5,616) & (\num{0.042}) & (\num{0.013}) & (\num{0.012}) & (\num{0.012}) & (\num{0.012}) & (\num{0.012}) & (\num{0.012})\\\cmidrule(lr){2-8}
\hspace{1em} Difference & \num{0.044} & \num{-0.024} & \num{-0.017} & \num{-0.020} & \num{-0.003} & \num{-0.028} & \num{0.003}\\
\hspace{2em} & (\num{0.060}) & (\num{0.018}) & (\num{0.018}) & (\num{0.017}) & (\num{0.018}) & (\num{0.018}) & (\num{0.018})\\\multicolumn{4}{l}{\textbf{Supports governing party}} \rule{0pt}{1.2\normalbaselineskip}\\
\hspace{1em} Not aligned & \num{0.113} & \num{-0.030} & \num{-0.027} & \num{-0.032} & \num{0.007} & \num{0.014} & \num{0.003}\\
\hspace{2em}(n = 7,360) & (\num{0.036}) & (\num{0.011}) & (\num{0.011}) & (\num{0.010}) & (\num{0.011}) & (\num{0.011}) & (\num{0.011})\\
\hspace{1em} Aligned & \num{-0.063} & \num{0.000} & \num{0.000} & \num{0.007} & \num{0.005} & \num{0.003} & \num{0.000}\\
\hspace{2em}(n = 3,171) & (\num{0.054}) & (\num{0.016}) & (\num{0.016}) & (\num{0.015}) & (\num{0.015}) & (\num{0.016}) & (\num{0.016})\\\cmidrule(lr){2-8}
\hspace{1em} Difference & \num{-0.175} & \num{0.030} & \num{0.027} & \num{0.039} & \num{-0.003} & \num{-0.011} & \num{-0.003}\\
\hspace{2em} & (\num{0.065}) & (\num{0.020}) & (\num{0.019}) & (\num{0.019}) & (\num{0.019}) & (\num{0.019}) & (\num{0.019})\\\multicolumn{4}{l}{\textbf{Digital literacy index}} \rule{0pt}{1.2\normalbaselineskip}\\
\hspace{1em} Below median & \num{0.068} & \num{-0.029} & \num{-0.028} & \num{-0.024} & \num{-0.006} & \num{0.007} & \num{-0.003}\\
\hspace{2em}(n = 5,418) & (\num{0.042}) & (\num{0.013}) & (\num{0.012}) & (\num{0.012}) & (\num{0.012}) & (\num{0.012}) & (\num{0.012})\\
\hspace{1em} Above median & \num{0.051} & \num{-0.013} & \num{-0.009} & \num{-0.017} & \num{0.020} & \num{0.015} & \num{0.008}\\
\hspace{2em}(n = 5,113) & (\num{0.042}) & (\num{0.013}) & (\num{0.013}) & (\num{0.012}) & (\num{0.013}) & (\num{0.014}) & (\num{0.014})\\\cmidrule(lr){2-8}
\hspace{1em}  Difference & \num{-0.017} & \num{0.016} & \num{0.020} & \num{0.007} & \num{0.026} & \num{0.009} & \num{0.011}\\
\hspace{2em} & (\num{0.060}) & (\num{0.018}) & (\num{0.018}) & (\num{0.017}) & (\num{0.018}) & (\num{0.018}) & (\num{0.018})\\\multicolumn{4}{l}{\textbf{Scientific knowledge index}} \rule{0pt}{1.2\normalbaselineskip}\\
\hspace{1em} Below median & \num{0.105} & \num{-0.037} & \num{-0.040} & \num{-0.032} & \num{-0.005} & \num{0.005} & \num{-0.010}\\
\hspace{2em}(n = 5,560) & (\num{0.041}) & (\num{0.013}) & (\num{0.013}) & (\num{0.012}) & (\num{0.013}) & (\num{0.013}) & (\num{0.013})\\
\hspace{1em} Above median & \num{0.010} & \num{-0.004} & \num{0.005} & \num{-0.007} & \num{0.020} & \num{0.017} & \num{0.015}\\
\hspace{2em}(n = 4,971) & (\num{0.043}) & (\num{0.013}) & (\num{0.013}) & (\num{0.012}) & (\num{0.012}) & (\num{0.013}) & (\num{0.013})\\\cmidrule(lr){2-8}
\hspace{1em} Difference & \num{-0.096} & \num{0.032} & \num{0.045} & \num{0.025} & \num{0.025} & \num{0.011} & \num{0.025}\\
\hspace{2em} & (\num{0.060}) & (\num{0.018}) & (\num{0.018}) & (\num{0.017}) & (\num{0.018}) & (\num{0.018}) & (\num{0.018})\\
\end{tabular}

}
   \caption{\footnotesize\textbf{Heterogeneity in treatment effects under averaged respondent-level treatments by selected covariates.} The sample is users in the evaluation stage, $n = 10,531$. 
   Columns denote response measures, which include discernment, a weighted sum of number of false sharing intentions (negatively weighted) and true sharing intentions (positively weighted); and for false and true posts separately, average propensity to share posts over any channel, over Messenger only, and on timeline only. 
   Estimates are of treatment effects averaged across the two respondent-level treatments, in contrast with the control condition. Estimates are produced from an augmented inverse probability weighted estimator, as described in \autoref*{section:estimation}, within specified subgroups. 
   }
  \refstepcounter{SItable}\label{tab:heterogeneity_treatment}
\end{table}

\begin{table}[H]
\small
   \centering
\resizebox{0.95\textwidth}{!}{ 

\begin{tabular}[t]{lccccccc}
 & \textbf{Sharing} &  & \textbf{False} &  &  & \textbf{True} & \\
 & \textbf{Discernment} & Any sharing & Messenger & Timeline & Any sharing & Messenger & Timeline\\\cmidrule(lr){2-2} \cmidrule(lr){3-5} \cmidrule(lr){6-8} \multicolumn{4}{l}{\textbf{Age}} \rule{0pt}{1.2\normalbaselineskip}\\
\hspace{1em} Below median & \num{0.019} & \num{-0.019} & \num{-0.020} & \num{-0.019} & \num{-0.017} & \num{-0.010} & \num{-0.021}\\
\hspace{2em}(n = 5,300) & (\num{0.043}) & (\num{0.014}) & (\num{0.013}) & (\num{0.013}) & (\num{0.013}) & (\num{0.014}) & (\num{0.014})\\
\hspace{1em} Above Median & \num{0.114} & \num{-0.027} & \num{-0.016} & \num{-0.032} & \num{0.034} & \num{0.033} & \num{0.025}\\
\hspace{2em}(n = 5,231) & (\num{0.048}) & (\num{0.014}) & (\num{0.014}) & (\num{0.013}) & (\num{0.013}) & (\num{0.014}) & (\num{0.014})\\\cmidrule(lr){2-8}
\hspace{1em} Difference & \num{0.096} & \num{-0.008} & \num{0.004} & \num{-0.012} & \num{0.051} & \num{0.043} & \num{0.046}\\
\hspace{2em} & (\num{0.064}) & (\num{0.020}) & (\num{0.019}) & (\num{0.018}) & (\num{0.019}) & (\num{0.020}) & (\num{0.019})\\\multicolumn{4}{l}{\textbf{Gender}} \rule{0pt}{1.2\normalbaselineskip}\\
\hspace{1em} Not male & \num{0.031} & \num{-0.007} & \num{-0.007} & \num{-0.009} & \num{0.009} & \num{0.030} & \num{-0.002}\\
\hspace{2em}(n = 4,915) & (\num{0.045}) & (\num{0.014}) & (\num{0.014}) & (\num{0.013}) & (\num{0.014}) & (\num{0.015}) & (\num{0.014})\\
\hspace{1em} Male & \num{0.097} & \num{-0.036} & \num{-0.027} & \num{-0.039} & \num{0.007} & \num{-0.006} & \num{0.005}\\
\hspace{2em}(n = 5,616) & (\num{0.045}) & (\num{0.014}) & (\num{0.013}) & (\num{0.013}) & (\num{0.012}) & (\num{0.013}) & (\num{0.013})\\\cmidrule(lr){2-8}
\hspace{1em} Difference & \num{0.067} & \num{-0.029} & \num{-0.029} & \num{-0.029} & \num{-0.002} & \num{-0.002} & \num{-0.002}\\
\hspace{2em} & (\num{0.064}) & (\num{0.020}) & (\num{0.020}) & (\num{0.020}) & (\num{0.019}) & (\num{0.019}) & (\num{0.019})\\\multicolumn{4}{l}{\textbf{Supports governing party}} \rule{0pt}{1.2\normalbaselineskip}\\
\hspace{1em} Not aligned & \num{0.116} & \num{-0.031} & \num{-0.028} & \num{-0.033} & \num{0.009} & \num{0.015} & \num{0.003}\\
\hspace{2em}(n = 7,360) & (\num{0.038}) & (\num{0.012}) & (\num{0.012}) & (\num{0.011}) & (\num{0.012}) & (\num{0.012}) & (\num{0.012})\\
\hspace{1em} Aligned & \num{-0.049} & \num{-0.004} & \num{0.005} & \num{-0.007} & \num{0.006} & \num{0.003} & \num{-0.001}\\
\hspace{2em}(n = 3,171) & (\num{0.058}) & (\num{0.017}) & (\num{0.017}) & (\num{0.016}) & (\num{0.016}) & (\num{0.017}) & (\num{0.017})\\\cmidrule(lr){2-8}
\hspace{1em} Difference & \num{-0.165} & \num{0.026} & \num{0.026} & \num{0.026} & \num{-0.003} & \num{-0.003} & \num{-0.003}\\
\hspace{2em} & (\num{0.069}) & (\num{0.021}) & (\num{0.021}) & (\num{0.021}) & (\num{0.020}) & (\num{0.020}) & (\num{0.020})\\\multicolumn{4}{l}{\textbf{Digital literacy index}} \rule{0pt}{1.2\normalbaselineskip}\\
\hspace{1em} Below median & \num{0.071} & \num{-0.031} & \num{-0.028} & \num{-0.030} & \num{-0.006} & \num{0.007} & \num{-0.004}\\
\hspace{2em}(n = 5,418) & (\num{0.045}) & (\num{0.014}) & (\num{0.013}) & (\num{0.013}) & (\num{0.013}) & (\num{0.013}) & (\num{0.013})\\
\hspace{1em} Above median & \num{0.062} & \num{-0.014} & \num{-0.008} & \num{-0.021} & \num{0.024} & \num{0.016} & \num{0.007}\\
\hspace{2em}(n = 5,113) & (\num{0.046}) & (\num{0.014}) & (\num{0.014}) & (\num{0.013}) & (\num{0.014}) & (\num{0.014}) & (\num{0.014})\\\cmidrule(lr){2-8}
\hspace{1em}  Difference & \num{-0.009} & \num{0.017} & \num{0.020} & \num{0.009} & \num{0.030} & \num{0.009} & \num{0.011}\\
\hspace{2em} & (\num{0.064}) & (\num{0.020}) & (\num{0.019}) & (\num{0.018}) & (\num{0.019}) & (\num{0.020}) & (\num{0.019})\\\multicolumn{4}{l}{\textbf{Scientific knowledge index}} \rule{0pt}{1.2\normalbaselineskip}\\
\hspace{1em} Below median & \num{0.139} & \num{-0.043} & \num{-0.045} & \num{-0.043} & \num{-0.003} & \num{0.007} & \num{-0.014}\\
\hspace{2em}(n = 5,560) & (\num{0.044}) & (\num{0.014}) & (\num{0.013}) & (\num{0.013}) & (\num{0.013}) & (\num{0.014}) & (\num{0.014})\\
\hspace{1em} Above median & \num{-0.015} & \num{0.000} & \num{0.012} & \num{-0.005} & \num{0.021} & \num{0.016} & \num{0.019}\\
\hspace{2em}(n = 4,971) & (\num{0.046}) & (\num{0.014}) & (\num{0.014}) & (\num{0.013}) & (\num{0.013}) & (\num{0.014}) & (\num{0.014})\\\cmidrule(lr){2-8}
\hspace{1em} Difference & \num{-0.154} & \num{0.043} & \num{0.057} & \num{0.038} & \num{0.024} & \num{0.009} & \num{0.033}\\
\hspace{2em} & (\num{0.064}) & (\num{0.020}) & (\num{0.019}) & (\num{0.018}) & (\num{0.019}) & (\num{0.020}) & (\num{0.019})\\
\end{tabular}

   }
      \caption{\textbf{Heterogeneity in treatment effects under accuracy nudge by selected covariates.} The sample is users in the evaluation stage, $n = 10,531$. 
   Columns denote response measures, described in the note to \autoref*{tab:heterogeneity_treatment}. 
   Estimates are of treatment effects under the accuracy nudge, in contrast with the control condition. Estimates are produced from an augmented inverse probability weighted estimator, as described in \autoref*{section:estimation}, within specified subgroups. 
   }
  \refstepcounter{SItable}\label{tab:heterogeneity_treatment_accuracy}
\end{table}

\begin{table}[H]
\small
   \centering
\resizebox{0.95\textwidth}{!}{ 

\begin{tabular}[t]{lccccccc}
 & \textbf{Sharing} &  & \textbf{False} &  &  & \textbf{True} & \\
 & \textbf{Discernment} & Any sharing & Messenger & Timeline & Any sharing & Messenger & Timeline\\\cmidrule(lr){2-2} \cmidrule(lr){3-5} \cmidrule(lr){6-8} \multicolumn{4}{l}{\textbf{Age}} \rule{0pt}{1.2\normalbaselineskip}\\
\hspace{1em} Below median & \num{-0.027} & \num{-0.004} & \num{-0.004} & \num{0.003} & \num{-0.018} & \num{-0.009} & \num{-0.019}\\
\hspace{2em}(n = 5,300) & (\num{0.047}) & (\num{0.015}) & (\num{0.015}) & (\num{0.014}) & (\num{0.015}) & (\num{0.015}) & (\num{0.015})\\
\hspace{1em} Above Median & \num{0.135} & \num{-0.036} & \num{-0.035} & \num{-0.034} & \num{0.028} & \num{0.030} & \num{0.024}\\
\hspace{2em}(n = 5,231) & (\num{0.053}) & (\num{0.016}) & (\num{0.015}) & (\num{0.015}) & (\num{0.015}) & (\num{0.015}) & (\num{0.015})\\\cmidrule(lr){2-8}
\hspace{1em} Difference & \num{0.161} & \num{-0.032} & \num{-0.031} & \num{-0.037} & \num{0.047} & \num{0.039} & \num{0.043}\\
\hspace{2em} & (\num{0.071}) & (\num{0.022}) & (\num{0.022}) & (\num{0.021}) & (\num{0.021}) & (\num{0.022}) & (\num{0.021})\\\multicolumn{4}{l}{\textbf{Gender}} \rule{0pt}{1.2\normalbaselineskip}\\
\hspace{1em} Not male & \num{0.042} & \num{-0.009} & \num{-0.012} & \num{-0.011} & \num{0.007} & \num{0.021} & \num{0.003}\\
\hspace{2em}(n = 4,915) & (\num{0.051}) & (\num{0.016}) & (\num{0.016}) & (\num{0.014}) & (\num{0.016}) & (\num{0.016}) & (\num{0.016})\\
\hspace{1em} Male & \num{0.063} & \num{-0.029} & \num{-0.026} & \num{-0.020} & \num{0.003} & \num{0.001} & \num{0.002}\\
\hspace{2em}(n = 5,616) & (\num{0.050}) & (\num{0.015}) & (\num{0.015}) & (\num{0.014}) & (\num{0.014}) & (\num{0.014}) & (\num{0.015})\\\cmidrule(lr){2-8}
\hspace{1em} Difference & \num{0.021} & \num{-0.020} & \num{-0.020} & \num{-0.020} & \num{-0.003} & \num{-0.003} & \num{-0.003}\\
\hspace{2em} & (\num{0.071}) & (\num{0.022}) & (\num{0.022}) & (\num{0.022}) & (\num{0.021}) & (\num{0.021}) & (\num{0.021})\\\multicolumn{4}{l}{\textbf{Supports governing party}} \rule{0pt}{1.2\normalbaselineskip}\\
\hspace{1em} Not aligned & \num{0.110} & \num{-0.030} & \num{-0.026} & \num{-0.032} & \num{0.006} & \num{0.013} & \num{0.003}\\
\hspace{2em}(n = 7,360) & (\num{0.043}) & (\num{0.013}) & (\num{0.013}) & (\num{0.012}) & (\num{0.013}) & (\num{0.013}) & (\num{0.013})\\
\hspace{1em} Aligned & \num{-0.076} & \num{0.003} & \num{-0.005} & \num{0.021} & \num{0.003} & \num{0.004} & \num{0.001}\\
\hspace{2em}(n = 3,171) & (\num{0.065}) & (\num{0.019}) & (\num{0.019}) & (\num{0.019}) & (\num{0.018}) & (\num{0.019}) & (\num{0.018})\\\cmidrule(lr){2-8}
\hspace{1em} Difference & \num{-0.186} & \num{0.033} & \num{0.033} & \num{0.033} & \num{-0.003} & \num{-0.003} & \num{-0.003}\\
\hspace{2em} & (\num{0.077}) & (\num{0.023}) & (\num{0.023}) & (\num{0.023}) & (\num{0.022}) & (\num{0.022}) & (\num{0.022})\\\multicolumn{4}{l}{\textbf{Digital literacy index}} \rule{0pt}{1.2\normalbaselineskip}\\
\hspace{1em} Below median & \num{0.066} & \num{-0.027} & \num{-0.029} & \num{-0.018} & \num{-0.006} & \num{0.006} & \num{-0.003}\\
\hspace{2em}(n = 5,418) & (\num{0.050}) & (\num{0.015}) & (\num{0.015}) & (\num{0.014}) & (\num{0.014}) & (\num{0.014}) & (\num{0.014})\\
\hspace{1em} Above median & \num{0.040} & \num{-0.012} & \num{-0.009} & \num{-0.014} & \num{0.016} & \num{0.014} & \num{0.008}\\
\hspace{2em}(n = 5,113) & (\num{0.051}) & (\num{0.016}) & (\num{0.016}) & (\num{0.015}) & (\num{0.016}) & (\num{0.016}) & (\num{0.016})\\\cmidrule(lr){2-8}
\hspace{1em}  Difference & \num{-0.026} & \num{0.016} & \num{0.020} & \num{0.004} & \num{0.022} & \num{0.008} & \num{0.011}\\
\hspace{2em} & (\num{0.071}) & (\num{0.022}) & (\num{0.022}) & (\num{0.021}) & (\num{0.021}) & (\num{0.022}) & (\num{0.022})\\\multicolumn{4}{l}{\textbf{Scientific knowledge index}} \rule{0pt}{1.2\normalbaselineskip}\\
\hspace{1em} Below median & \num{0.071} & \num{-0.030} & \num{-0.035} & \num{-0.022} & \num{-0.008} & \num{0.004} & \num{-0.005}\\
\hspace{2em}(n = 5,560) & (\num{0.050}) & (\num{0.015}) & (\num{0.015}) & (\num{0.014}) & (\num{0.015}) & (\num{0.015}) & (\num{0.015})\\
\hspace{1em} Above median & \num{0.034} & \num{-0.008} & \num{-0.003} & \num{-0.009} & \num{0.019} & \num{0.017} & \num{0.011}\\
\hspace{2em}(n = 4,971) & (\num{0.051}) & (\num{0.016}) & (\num{0.015}) & (\num{0.015}) & (\num{0.014}) & (\num{0.015}) & (\num{0.015})\\\cmidrule(lr){2-8}
\hspace{1em} Difference & \num{-0.037} & \num{0.022} & \num{0.032} & \num{0.012} & \num{0.027} & \num{0.013} & \num{0.017}\\
\hspace{2em} & (\num{0.071}) & (\num{0.022}) & (\num{0.022}) & (\num{0.021}) & (\num{0.021}) & (\num{0.022}) & (\num{0.021})\\
\end{tabular}

   }
   \caption{\textbf{Heterogeneity in treatment effects under Facebook tips by selected covariates.} The sample is users in the evaluation stage, $n = 10,531$. 
   Columns denote response measures, described in the note to \autoref*{tab:heterogeneity_treatment}. 
   Estimates are of treatment effects under the Facebook tips, in contrast with the control condition. Estimates are produced from an augmented inverse probability weighted estimator, as described in \autoref*{section:estimation}, within specified subgroups. 
   }
  \refstepcounter{SItable}\label{tab:heterogeneity_treatment_facebook}
\end{table}

\clearpage

\end{document}